\RequirePackage{ifpdf}
\ifpdf 
\documentclass[pdftex]{sigma}
\else
\documentclass{sigma}
\fi

\begin{document}
\allowdisplaybreaks

\def\1{\'{\i}}
\def\k{\kappa}
\def\Sk{{\rm\ \!S}}
\def\Ck{{\rm\ \!C}}
\def\Tk{{\rm\ \!T}}

\renewcommand{\PaperNumber}{010}

\FirstPageHeading

\ShortArticleName{Superintegrability on 3D Spaces  of Constant Curvature}

\ArticleName{Superintegrability on Three-Dimensional Riemannian  and Relativistic
Spaces of Constant Curvature}

\Author{Francisco Jos\'e HERRANZ~$^\dag$ and \'Angel BALLESTEROS~$^\ddag$}
\AuthorNameForHeading{F.J. Herranz and \'A. Ballesteros}

\Address{$^\dag$~Departamento de F\1sica, Escuela Polit\'ecnica Superior, Universidad de Burgos,\\
$\phantom{^\dag}{}$~09001 Burgos, Spain}

\EmailD{\href{mailto:fjherranz@ubu.es}{fjherranz@ubu.es}}

\Address{$^\ddag$~Departamento de F\1sica, Facultad de Ciencias, Universidad de Burgos,\\
$\phantom{^\ddag}{}$~09001 Burgos, Spain}
\EmailD{\href{mailto:angelb@ubu.es}{angelb@ubu.es}}

\ArticleDates{Received December 21, 2005, in final form January 20,
2006; Published online January 24, 2006}

\Abstract{A family of classical superintegrable Hamiltonians,    depending on an
arbitrary radial function,  which are defined on the 3D
spherical, Euclidean and hyperbolic spaces as well as on the  (2+1)D anti-de
Sitter, Minkowskian and de Sitter spacetimes is constructed.
Such systems
admit three integrals of the motion (besides the Hamiltonian)  which are
explicitly given in terms of ambient and geodesic polar coordinates. The
resulting expressions cover the six spaces in a unified way as these are
parametrized by two contraction parameters that govern the curvature and
the signature of the metric on each space. Next two maximally
superintegrable Hamiltonians are identified within the initial
superintegrable family by finding the remaining  constant of the
motion. The former potential is
the superposition of a (curved) central  harmonic oscillator   with
 other three oscillators or centrifugal barriers (depending on each
specific space), so that this generalizes the Smorodinsky--Winternitz system. The
latter one is a superposition of the Kepler--Coulomb potential with another two oscillators
or centrifugal barriers. As a byproduct, the
Laplace--Runge--Lenz vector for these spaces is deduced.  Furthermore both  potentials  are
  analysed in detail for each particular space.
 Some comments on their generalization to arbitrary dimension
 are also presented.}


\Classification{37J35; 22E60; 37J15; 70H06}

\renewcommand{\theequation}{\arabic{section}.\arabic{equation}}

\section{Introduction}

In~\cite{6} Evans obtained a classification of classical  superintegrable systems~\cite{Per}
on the three-dimensional (3D) Euclidean space ${\bf E}^3$. At this dimension   he called
{\it minimally} superintegrable systems  those endowed with {\it three} constants of the
motion besides the Hamiltonian, that is, they have one  constant more  than those necessary  to
ensure complete integrability, but one less than the necessary
number to determine maximal superintegrability.   Amongst
the resulting potentials let us consider
\begin{gather}
{\cal U}={\cal F}(r)+\frac{\beta_1}{x^2}+\frac{\beta_2}{y^2}+\frac{\beta_3}{z^2},
\label{aa}
\end{gather}
where ${\cal F}(r)$ is an arbitrary  smooth function, the three $\beta_i$ are arbitrary real
parameters, $(x,y,z)$ are Cartesian coordinates on ${\bf E}^3$, and $r=\sqrt{x^2+y^2+z^2}$.
Thus this potential is formed by a central term with three centrifugal barriers. Next by
analysing the radial function ${\cal F}(r)$ two relevant and well known expressions arise
  conveying the additional constant of the motion. These two cases then appear in the
classification by Evans as {\it maximally} superintegrable systems as they have the maximum
number of functionally independent constants of the motion, {\it four} ones plus the
Hamiltonian. Explicitly, these are:
\begin{itemize}\vspace{-2mm}
\itemsep=0pt
\item   The
Smorodinsky--Winternitz (SW) potential~\cite{9} when ${\cal F}(r)=\beta_0 r^2$:
 \begin{gather}
{\cal U}^{\rm SW}=\beta_0
\left(x^2+y^2+z^2\right)+\frac{\beta_1}{x^2}+\frac{\beta_2}{y^2}+\frac{\beta_3}{z^2},
\label{ab}
\end{gather}
which corresponds to the superposition of a harmonic oscillator with angular frequency
$\sqrt{\beta_0}$ and the three centrifugal terms.

\item And a generalized Kepler--Coulomb (GKC) potential when ${\cal F}(r)=-k/r$:
 \begin{gather}
{\cal U}^{\rm GKC}=-\frac{k}{
\sqrt{x^2+y^2+z^2}}+\frac{\beta_1}{x^2}+\frac{\beta_2}{y^2} ,
\label{ac}
\end{gather}
which is formed by   the proper Kepler--Coulomb (KC) potential with parameter $k$ together
with {\it two} of the famous centrifugal  terms.
\vspace{-2mm}

\end{itemize}

Superintegrable systems on ${\bf E}^2$ and ${\bf E}^3$~\cite{6,25} have also been  implemented
on the two   classical Riemannian spaces of constant curvature. In particular, some
superintegrable systems on the 2D and 3D spheres, ${\bf S}^2$ and ${\bf S}^3$, have been
studied in~\cite{11}, on the hyperbolic plane~${\bf H}^2$ in~\cite{19,20}, while on ${\bf
H}^3$ can be found in~\cite{12}. Moreover classifications of superintegrable systems on ${\bf
S}^2$ and~${\bf H}^2$  have been carried out in~\cite{18,21,24,27}.
These results contain the corresponding (curved) harmonic oscillator~\cite{Higgs,26} and
KC potential~\cite{Schrodinger}, which in arbitrary dimension correspond, in this order, to
the following radial potential
\begin{gather}
{\cal F}(r)=\left\{
\begin{array}{ll}
 \beta_0 \tan^2 r ,&   {\rm on}\quad {\bf S}^N; \\
 \beta_0\, r^2 , &
  {\rm on}\quad {\bf E}^N ;\\
 \beta_0 \tanh^2 r ,&  {\rm on}\quad {\bf H}^N.
\end{array}\right.
\qquad
{\cal F}(r)=\left\{
\begin{array}{ll}
-k/ \tan r ,&   {\rm on}\quad {\bf S}^N; \\
-k/ r , &
  {\rm on}\quad {\bf E}^N ;\\
- k/ \tanh r ,&  {\rm on}\quad {\bf H}^N.
\end{array}\right.
\label{ad}
\end{gather}
We recall that the SW system on ${\bf S}^N$ and ${\bf H}^N$ have been constructed
in~\cite{VulpiLett,CRMVulpi} (curved harmonic oscillator plus $N$ terms) showing that this
keeps  maximal superintegrability for any value of the curvature.

However, as far as we know, the construction of the GKC potential on ${\bf S}^N$ and ${\bf
H}^N$ as well as which are the corresponding SW and GKC systems on the relativistic spacetimes
of constant curvature is still lacking, that is, also covering the  anti-de Sitter,
Minkowskian and de Sitter spacetimes.  The aim of this paper is to present all of these
Hamiltonians on these {\it six} 3D spaces in a unified setting by making use of two explicit
contraction parameters which determine the curvature and the signature of the metric. In this
sense, the results here presented
  can be considered as the  cornerstone for a further
generalization of all of these systems to arbitrary dimension. In this respect, we would like
to mention that although very recently such potentials have been deduced on the $(1+1)$D
relativistic spacetimes~\cite{jpa2D,car2}, this low dimension does not show the guide for  a
direct generalization to $N$D.

The structure of this paper is as follows. The next section
contains the necessary basics on the  Lie groups of
isometries on the six spaces together with the two coordinate systems we shall deal with
throughout the paper:   ambient (Weierstrass) coordinates in an auxiliary linear
space~$\mathbb R^{4}$ and   intrinsic geodesic polar (spherical) coordinates.
The kinetic energy determining the geodesic motion is then studied in Section~3
by starting from the metric.
The   generalization of the Euclidean family (\ref{aa}) to these spaces  is
developed in Section~4 in such a manner that
general and global expressions for the Hamiltonian and its
three integrals of motion are explicitly given.

The next two sections are devoted to the
study of two maximal superintegrable Hamiltonians arising in the above family by choosing in
an adequate way the radial function ${\cal F}(r)$ (fulfil\-ling~(\ref{ad}) for the Riemannian
spaces) and  finding at the same time the remaining constant of the motion.
In this way we obtain the generalization of the SW (\ref{ab}) and GKC (\ref{ac}) potentials   for any value of the curvature and signature of the
metric. Furthermore a detail description of such systems is performed on each particular
space. We stress that, by following the geometrical interpretation formerly introduced
in~\cite{ran,ran1,ran2} and generalized in~\cite{jpa2D,VulpiLett,CRMVulpi},  the SW potential
is interpreted as the superposition of a central harmonic oscillator with three non-central
oscillators or centrifugal barriers according to each specific space.
 Likewise, the   GKC system can be seen as the superposition of the KC potential with two
oscillators or centrifugal barriers; in this case, we moreover deduce the corresponding
Laplace--Runge--Lenz vector. Finally, some remarks and
comments mainly concerning the pattern for the construction of such systems for arbitrary
dimension close the paper.

\setcounter{equation}{0}

\section{Riemannian spaces and relativistic spacetimes}

Let us consider a subset of real Lie algebras contained in the family of the  Cayley--Klein
ortho\-go\-nal   algebras~\cite{CK3,Groma}. These can also be obtained
as the   $\mathbb Z_2\otimes
\mathbb Z_2$ graded contractions of $so(4)$ and are denoted $so_{\k_1,\k_2}(4)$ where $\k_1$
and $\k_2$ are two real contraction parameters. The   Lie brackets of
$so_{\k_1,\k_2}(4)$ in the basis spanned by
$\{J_{\mu\nu}\}$ where $\mu,\nu=0,1,2,3$ and $\mu<\nu$ read~\cite{CK3}
\begin{alignat}{5}
&[J_{12},J_{13}]=\k_2 J_{23}, \quad && [J_{12},J_{23}]=-J_{13},&& [J_{13},J_{23}]=J_{12}, & \nonumber\\
&[J_{12},J_{01}]=  J_{02},&& [J_{13},J_{01}]=J_{03},&& [J_{23},J_{02}]=J_{03}, & \nonumber\\
&[J_{12},J_{02}]=-\k_2 J_{01},\qquad && [J_{13},J_{03}]=-\k_2 J_{01},\qquad &&  [J_{23},J_{03}]=-J_{02},& \nonumber\\
& [J_{01},J_{02}]= \k_1 J_{12},&& [J_{01},J_{03}]= \k_1 J_{13},&& [J_{02},J_{03}]=\k_1\k_2 J_{23},\nonumber\\
& [J_{01},J_{23}]=0,&& [J_{02},J_{13}]= 0,&&  [J_{03},J_{12}]=0.& \label{ba}
\end{alignat}
There are two Casimir invariants
\begin{gather}
{\cal C}_1= \k_2 J_{01}^2+J_{02}^2+J_{03}^2+\k_1J_{12}^2+\k_1J_{13}^2+\k_1\k_2J_{23}^2,\nonumber\\
{\cal C}_2=\k_2 J_{01}J_{23}-J_{02}J_{13}+J_{03}J_{12} ,
\label{bba}
\end{gather}
where ${\cal C}_1$ is associated to the Killing--Cartan form.

Let us explain
the geometrical role of the contraction parameters $\k_1$ and $\k_2$. The involutive
automorphisms defined by
\begin{gather*}
\Theta_0: \ \ J_{ij}\to J_{ij},\quad J_{0i}\to -J_{0i},\quad i=1,2,3,\\
\Theta_{01}: \ \{J_{01},J_{23}\}\to  \{J_{01},J_{23}\},\quad  \{J_{0j},J_{1j}\}\to
-\{J_{0j},J_{1j}\},\quad j=2,3,
\end{gather*}
generate a $\mathbb Z_2\otimes
\mathbb Z_2$-grading of $so_{\k_1,\k_2}(4)$ in such a manner that $\k_1$ and $\k_2$ are two
graded contraction parameters coming from the $\mathbb Z_2$-grading determined by $\Theta_0$
and $\Theta_{01}$, respectively.
By scaling the Lie generators each parameter $\k_i$ can be reduced to either $+1$, 0 or $-1$;
the vanishment of $\k_i$ is   equivalent to apply an In\"on\"u--Wigner contraction.

Furthermore, these automorphisms
 give rise to the following Cartan
decompositions:
\begin{alignat*}{5}
& so_{\k_1,\k_2}(4)={\mathfrak{h}_0}\oplus  {\mathfrak{p}_0},&&
{\mathfrak{h}_0}=\langle J_{12},J_{13},J_{23}\rangle=so_{\k_2}(3),&&
{\mathfrak{p}_0}=\langle J_{01},J_{02},J_{03}\rangle,& \\
& so_{\k_1,\k_2}(4)={\mathfrak{h}_{01}}\oplus  {\mathfrak{p}_{01}},\qquad &&
{\mathfrak{h}_{01}}=\langle J_{01},J_{23}\rangle=so_{\k_1}(2)\oplus so(2),\qquad &&
{\mathfrak{p}_{01}}=\langle J_{02},J_{03},J_{12},J_{13}\rangle .&
\end{alignat*}
If $H_0$ and $H_{01}$ denote the Lie subgroups with Lie algebras ${\mathfrak{h}_0}$ and
${\mathfrak{h}_{01}}$,   we obtain two families of  symmetric homogeneous
spaces~\cite{13}, namely the usual 3D space of
points  $SO_{\k_1,\k_2}(4)/H_0$ and the 4D space of lines  $SO_{\k_1,\k_2}(4)/H_{01}$, which
have constant curvature equal to $\k_1$ and $\k_2$, respectively.

We shall make use of the
former space which has a metric with a   signature
governed by $\k_2$ as ${\rm diag}(+1,\k_2,\k_2)$ and we denote it
\[
{\mathbb S}^3_{[\k_1]\k_2}=SO_{\k_1,\k_2}(4)/SO_{\k_2}(3).
\]
 Thus when $\k_2$ is positive we recover the
three classical Riemannian spaces, while if this is negative we find a Lorentzian metric.
In this case, there is a kinematical interpretation for the homogeneous spaces.
Let $P_0$, $P_i$, $K_i$ and $J$ $(i=1,2)$ the usual generators of time
translation, space translations, boosts and spatial rotations,
respectively. Under the following identification
\begin{gather}
P_0=J_{01}, \qquad  P_i=J_{0\, i+1} , \qquad
 K_i=J_{1\, i+1}, \qquad J=J_{23}, \qquad i=1,2,
\label{be}
\end{gather}
the three   algebras   with
$\k_2=-1/c^2< 0$ ($c$ is the speed of light) are the Lie algebras  of the groups of motions
of
$(2+1)$D relativistic spacetime models. Thus the commutation relations~(\ref{ba})  read now
\begin{alignat}{7}
& [J,K_i]=\epsilon_{ij}K_j,\qquad &&  [K_1,K_2]=-\frac {1}{c^2}\, J,\qquad &&
 [P_0,K_i]=-P_i,\qquad  && [P_i,K_j]=-\frac {1}{c^2}\,\delta_{ij}P_0, & \nonumber\\
& [J,P_i]=\epsilon_{ij}P_j ,\qquad  && [P_1,P_2]=-\frac {\k_1}{c^2}\,J,\qquad &&
[P_0,P_i]=\k_1 K_i,\qquad  && [P_0,J_i]=0, & \label{bf}
\end{alignat}
where $\epsilon_{ij}$ is a skew-symmetric tensor such that $\epsilon_{12}=1$.
In this framework the curvature of the spacetime can be written in terms of the (time) universe
radius $\tau$ as
$\k_1=\pm 1/\tau^2$ (which is also proportional to the cosmological constant). The Casimir
invariants (\ref{bba}), ${\cal C}_1$  and  ${\cal C}_2$,  correspond to the energy and
angular momentum of a particle in the free kinematics of the relativistic spacetime:
\begin{gather}
 {\cal C}_1= -\frac 1{c^2}\, P_0^2+P_1^2+P_2^2+\k_1\left ( K_1^2+K_2^2\right)
-\frac{\k_1}{c^2}\,J^2,\nonumber\\
 {\cal C}_2=-\frac 1{c^2}\, P_0 J-P_1 K_2 +P_2 K_1.
\label{bg}
\end{gather}

On the other hand, if $\k_2=0$ we obtain a degenerate metric which corresponds to Newtonian
spacetimes. Since our aim is to construct superintegrable systems on these homogeneous spaces,
for which the kinetic energy is provided by the metric, we avoid the contraction $\k_2=0$.
The resulting {\it six} particular spaces contained in the
family  ${\mathbb S}^3_{[\k_1]\k_2}$ are  displayed in Table~1.

\bigskip
{\footnotesize

 \noindent
{\bf Table 1.} 3D symmetric homogeneous spaces ${\mathbb
S}^3_{[\k_1]\k_2}=SO_{\k_1,\k_2}(4)/SO_{\k_2}(3)$ and their metric in geodesic polar
coor\-dinates according to   $\k_1\in\{+1,0,-1\}$ and $\k_2\in\{+1,-1\}$.}
\begin{center}\footnotesize
$\begin{array}{ll}
\hline
\\[-6pt]
{\mbox {3D Riemannian spaces}}&\quad{\mbox  {$(2+1)$D
 Relativistic spacetimes}}\\[4pt]
\hline
\\[-6pt]
\mbox {$\bullet$ Spherical space  ${\bf S}^3$}&\quad\mbox {$\bullet$ Anti-de Sitter
spacetime ${\bf AdS}^{2+1}$}\\[4pt]
 \displaystyle{{\mathbb S}^3_{[+]+}=SO(4)/SO(3) } &\quad
 \displaystyle{{\mathbb S}^3_{[+]-}=SO(2,2)/SO(2,1) } \\[8pt]
 \displaystyle{{\rm d} s^2=
 {\rm d} r^2+   \sin^2 r\, {\rm d} \theta^2 +\sin^2 r\sin^2\theta \,{\rm d}
\phi^2   } &\quad
 \displaystyle{{\rm d} s^2=
 {\rm d} r^2-\sin^2 r\, {\rm d} \theta^2 -\sin^2 r \sinh^2\theta\, {\rm d}
\phi^2  } \\[8pt]
\mbox {$\bullet$ Euclidean space  ${\bf E}^3$}&\quad\mbox {$\bullet$ Minkowskian spacetime ${\bf
M}^{2+1}$}\\[4pt]
  \displaystyle{{\mathbb S}^3_{[0]+}=ISO(3)/SO(3) } &\quad
 \displaystyle{{\mathbb S}^3_{[0]-}=ISO(2,1)/SO(2,1) } \\[8pt]
 \displaystyle{{\rm d} s^2=
 {\rm d} r^2+     r^2\, {\rm d} \theta^2 +  r^2\sin^2\theta \,{\rm d}
\phi^2   } &\quad
 \displaystyle{{\rm d} s^2=
 {\rm d} r^2-r^2\, {\rm d} \theta^2 -r^2 \sinh^2\theta\, {\rm d}
\phi^2  } \\[8pt]
\mbox {$\bullet$  Hyperbolic space ${\bf H}^3$}&\quad\mbox {$\bullet$  De Sitter
spacetime ${\bf dS}^{2+1}$}\\[4pt]
  \displaystyle{{\mathbb S}^3_{[-]+}=SO(3,1)/SO(3) } &\quad
 \displaystyle{{\mathbb S}^3_{[-]-}=SO(3,1)/SO(2,1) } \\[8pt]
\displaystyle{{\rm d} s^2=
 {\rm d} r^2+   \sinh^2 r\, {\rm d} \theta^2 +\sinh^2 r\sin^2\theta \,{\rm d}
\phi^2   } &\quad
 \displaystyle{{\rm d} s^2=
 {\rm d} r^2-\sinh^2 r\, {\rm d} \theta^2 -\sinh^2 r \sinh^2\theta\, {\rm d}
\phi^2  } \\[8pt]
\hline
\end{array}
$\end{center}

\subsection{Vector model and ambient coordinates}

The vector representation of $so_{\k_1,\k_2}(4)$ is  given by the following $4\times 4$ real
matrices~\cite{CK3}:
\begin{alignat}{5}
& J_{01}=\left(\begin{array}{cccc}
\cdot& -\k_1 & \cdot & \cdot\cr
1& \cdot & \cdot & \cdot\cr
\cdot& \cdot & \cdot & \cdot\cr
\cdot& \cdot & \cdot & \cdot
\end{array}\right) ,\qquad  &&  J_{12}=\left(\begin{array}{cccc}
\cdot& \cdot & \cdot & \cdot\cr
\cdot& \cdot & -\k_2 & \cdot\cr
\cdot& 1 & \cdot & \cdot\cr
\cdot& \cdot & \cdot & \cdot
\end{array}\right) ,& \nonumber\\
& J_{02}=\left(\begin{array}{cccc}
\cdot& \cdot & -\k_1\k_2 & \cdot\cr
\cdot& \cdot & \cdot & \cdot\cr
1& \cdot & \cdot & \cdot\cr
\cdot& \cdot & \cdot & \cdot
\end{array}\right) ,\qquad &&
 J_{13}=\left(\begin{array}{cccc}
\cdot& \cdot & \cdot & \cdot\cr
\cdot& \cdot & \cdot & -\k_2\cr
\cdot& \cdot & \cdot & \cdot\cr
\cdot& 1 & \cdot & \cdot
\end{array}\right) ,& \nonumber\\
& J_{03}=\left(\begin{array}{cccc}
\cdot& \cdot & \cdot & -\k_1\k_2\cr
\cdot& \cdot & \cdot & \cdot\cr
\cdot& \cdot & \cdot & \cdot\cr
1& \cdot & \cdot & \cdot
\end{array}\right),\qquad &&
 J_{23}=\left(\begin{array}{cccc}
\cdot& \cdot & \cdot & \cdot\cr
\cdot& \cdot & \cdot & \cdot\cr
\cdot& \cdot & \cdot & -1\cr
\cdot& \cdot & 1 & \cdot
\end{array}\right) .
& \label{ca}
\end{alignat}
Their exponential provides the corresponding one-parametric subgroups of
$SO_{\k_1,\k_2}(4)$:
\begin{gather}
{\rm e}^{xJ_{01}}=\left(\begin{array}{cccc}\!\!
\Ck_{\k_1}(x)& -\k_1 \Sk_{\k_1}(x)& \cdot & \cdot\cr
\Sk_{\k_1}(x)& \Ck_{\k_1}(x) & \cdot & \cdot\cr
\cdot& \cdot & 1 & \cdot\cr
\cdot& \cdot & \cdot & 1
\end{array}\!\!\right) ,\qquad  {\rm e}^{xJ_{12}}=\left(\begin{array}{cccc}\!\!
1& \cdot & \cdot & \cdot\cr
\cdot& \Ck_{\k_2}(x) & -\k_2 \Sk_{\k_2}(x)& \cdot\cr
\cdot& \Sk_{\k_2}(x) & \Ck_{\k_2}(x) & \cdot\cr
\cdot& \cdot & \cdot & 1
\end{array}\!\!\right) ,\nonumber\\
{\rm e}^{xJ_{02}}=\left(\begin{array}{cccc}
\!\! \Ck_{\k_1\k_2}(x)& \cdot &\!\!\! -\k_1\k_2 \Sk_{\k_1\k_2}(x)&\!\! \cdot\cr
\!\!\!\cdot& 1 & \cdot &\!\! \cdot\cr
\!\!\!\Sk_{\k_1\k_2}(x)& \cdot & \Ck_{\k_1\k_2}(x) &\!\! \cdot\cr
\!\!\!\cdot& \cdot & \cdot &\!\! 1
\end{array}\!\!\right) ,\qquad
{\rm e}^{x J_{13}}=\left(\begin{array}{cccc}\!\!
1& \cdot & \cdot & \cdot\cr
\cdot& \!\!  \Ck_{\k_2}(x) & \cdot &\!\! -\k_2 \Sk_{\k_2}(x)\cr
\cdot& \cdot & 1 & \cdot\cr
\cdot&\!\! \Sk_{\k_2}(x) & \cdot &\!\! \Ck_{\k_2}(x)
\end{array}\!\!\right) ,\nonumber\\
{\rm e}^{x J_{03}}=\left(\!\!\!\begin{array}{cccc}
\Ck_{\k_1\k_2}(x)& \cdot & \cdot & -\k_1\k_2 \Sk_{\k_1\k_2}(x)\cr
\cdot& 1 & \cdot & \cdot\cr
\cdot& \cdot & 1 & \cdot\cr
\Sk_{\k_1\k_2}(x)& \cdot & \cdot & \Ck_{\k_1\k_2}(x)
\end{array}\!\!\right),\qquad
{\rm e}^{x J_{23}}=\left(\!\!\begin{array}{cccc}
1& \cdot & \cdot & \cdot\cr
\cdot& 1 & \cdot & \cdot\cr
\cdot& \cdot & \!\cos x\! & \!-\sin x\cr
\cdot& \cdot & \!\sin x \! & \cos x
\end{array}\!\!\right) ,\!\!\!
\label{cb}
\end{gather}
where we have introduced the $\k$-dependent cosine and sine functions defined
by~\cite{CK2,PL}
\begin{gather}
\Ck_{\k}(x) =\sum_{l=0}^{\infty}(-\k)^l\frac{x^{2l}}
{(2l)!}=\left\{
\begin{array}{ll}
\displaystyle  \cos {\sqrt{\k}\, x} ,&\quad  \k>0; \vspace{1mm}\\
\qquad 1 , &\quad
  \k=0 ;\vspace{1mm}\\
\displaystyle \cosh {\sqrt{-\k}\, x} ,&\quad   \k<0 ,
\end{array}\right.
\nonumber
\\
   \Sk{_\k}(x) =\sum_{l=0}^{\infty}(-\k)^l\frac{x^{2l+1}}{ (2l+1)!}
= \left\{
\begin{array}{ll}
  \tfrac{1}{\sqrt{\k}} \sin {\sqrt{\k}\, x} ,&\quad  \k>0 ;\vspace{1mm}\\
\qquad x , &\quad
  \k=0 ;\vspace{1mm}\\
\tfrac{1}{\sqrt{-\k}} \sinh {\sqrt{-\k}\, x} ,&\quad  \k<0 .
\end{array}\right.
\nonumber
\end{gather}
Notice that $\k\in\{\k_1,\k_1\k_2,\k_2\}$. The tangent is defined
as $\Tk_\k(x)=\Sk_\k(x)/\Ck_\k(x)$.
 Properties and  trigonometric  relations  for these $\k$-functions, which are necessary
in the  further computations, can be found in~\cite{trigo,conf}; for instance,
\[
\Ck^2_\k(x)+\k\,\Sk^2_\k(x)=1,\qquad \frac{ {\rm d}}
{{\rm d} x}\Ck_\k(x)=-\k\,\Sk_\k(x),\qquad
\frac{ {\rm d}}
{{\rm d} x}\Sk_\k(x)= \Ck_\k(x) .
\]

Under the above matrix algebra and group representations  it is verified that
\[
X^T \mathbb I_{\k}+\mathbb I_{\k} X=0,\quad  X\in so_{\k_1,\k_2}(4),\qquad
Y^T \mathbb I_{\k} Y=\mathbb I_{\k} ,\quad  Y\in SO_{\k_1,\k_2}(4),
\]
($X^T$ is the transpose
matrix of $X$) with respect to the bilinear form
\[
\mathbb I_{\k}={\rm diag}(+1,\k_1,\k_1\k_2,\k_1\k_2) .
\]
 Therefore
$SO_{\k_1,\k_2}(4)$ is a group of isometries of  $\mathbb I_{\k}$ acting on
a linear ambient space
$\mathbb R^4=(x_0,x_1,x_2,x_3)$ through matrix multiplication. The origin $O$ in ${\mathbb
S}^3_{[\k_1]\k_2}$ has ambient coordinates $O =(1,0,0,0)$ and this point is invariant under
the subgroup $H_0=SO_{\k_2}(3)=\langle J_{12},J_{13},J_{23}\rangle$ (see~(\ref{cb})). The orbit
of $O$ corresponds to the homogeneous space ${\mathbb S}^3_{[\k_1]\k_2}$ which is contained
in the ``sphere''
\begin{gather}
\Sigma\equiv x_0^2+\k_1 x_1^2+\k_1\k_2 x_2^2 +\k_1\k_2 x_3^2=1 ,
\label{cd}
\end{gather}
determined by $\mathbb I_{\k}$. The ambient coordinates $(x_0,x_1,x_2,x_3)$, subjected to
(\ref{cd}),  are also called {\it Weierstrass coordinates}.  The metric on ${\mathbb
S}^3_{[\k_1]\k_2}$
follows from    the flat ambient metric in $\mathbb R^{4}$ divided by the curvature and
restricted to $\Sigma$:
\begin{gather}
{\rm d} s^2=\frac {1}{\k_1}
\left({\rm d} x_0^2+   \k_1 {\rm d} x_1^2+\k_1\k_2 {\rm d} x_2^2 +\k_1\k_2 {\rm d} x_3^2
\right)\Big|_{\Sigma}.
\label{ce}
\end{gather}
A differential realization of $so_{\k_1,\k_2}(4)$, fulfilling (\ref{ba}), as first-order vector
fields in the ambient coordinates is provided by the vector representation (\ref{ca}) and reads
\begin{alignat}{3}
& J_{01}=\k_1 x_1\partial_0 -x_0\partial_1,\qquad &&
J_{0j}=\k_1\k_2 x_j\partial_0 -x_0\partial_j,&\nonumber\\
& J_{23}=  x_3\partial_2 -x_2\partial_3, \quad && J_{1j}=\k_2 x_j\partial_1 -x_1\partial_j, &
\label{cf}
\end{alignat}
where $j=2,3$ and $\partial_\mu=\partial/\partial x_\mu$.

\subsection{Geodesic polar coordinate system}

Let us consider a point $Q$ in ${\mathbb
S}^3_{[\k_1]\k_2}$ with Weierstrass coordinates $(x_0,x_1,x_2,x_3)$. This can be
parametrized in terms of three intrinsic quantities of the space itself in different
ways. We shall make use of the {\it geodesic polar coordinates} $(r,\theta,\phi)$ which are
defined through the following action of the one-parametric subgroups (\ref{cb}) on the
origin
$O=(1,0,0,0)$:
\begin{gather}
Q(r,\theta,\phi)=\exp\{\phi J_{23}\}\exp\{\theta J_{12}\}\exp\{r
J_{01}\}O,\nonumber\\
\left(\begin{array}{c}
x_0\cr
x_1\cr
x_2\cr
x_3
\end{array}\right)=
\left(\begin{array}{c}
\Ck_{\k_1}(r)\cr
\Sk_{\k_1}(r)\Ck_{\k_2}(\theta)\cr
\Sk_{\k_1}(r)\Sk_{\k_2}(\theta)\cos\phi\cr
\Sk_{\k_1}(r)\Sk_{\k_2}(\theta)\sin\phi
\end{array}\right).
\label{da}
\end{gather}

Let   $l_1$  be a (time-like) geodesic  and $l_2$, $l_3$ two other (space-like) geodesics
in ${\mathbb S}^3_{[\k_1]\k_2}$ orthogonal at
$O$    in such a manner that each translation generator $J_{0i}$ moves the origin along
$l_i$. Then the (physical) geometrical meaning of the coordinates $(r,\theta,\phi)$ is as
follows.
\begin{itemize}\vspace{-2mm}
\itemsep=0pt
\item The radial coordinate $r$ is the distance between $Q$ and   $O$
measured along the  (time-like)  geodesic $l$ that joins both points. In the curved
Riemannian spaces with
$\k_1=\pm 1/R^2$,
$r$~has dimensions of {\it length}, $[r]=[R]$; notice however that   the
dimensionless  coordina\-te~$r/R$  is usually
taken  instead of $r$, and so the former is considered as an ordinary angle (see,
e.g.,~\cite{17}). In the relativistic spacetimes with $\k_1=\pm 1/\tau^2$,  $r$ has
dimensions of a~time-like length, that is,
$[r]=[\tau]$.

\item The coordinate  $\theta$ is an ordinary angle  in the three Riemannian spaces
($\k_2=+1$) and this parametrizes the orientation of $l$ with respect to $l_1$, whilst $\theta$
corresponds to a rapidity in the spacetimes ($\k_2=-1/c^2$)  with dimensions
$[\theta]=[c]$.

\item Finally, $\phi$ is an ordinary   angle for the six spaces that determines
the orientation of $l$ with respect to the reference flag spanned by $l_1$ and $l_2$, that
is, the 2-plane $l_1l_2$.
\vspace{-2mm}

\end{itemize}

In the Riemannian spaces $(r,\theta,\phi)$ parametrize the complete space, while in the
spacetimes these only cover the time-like region (in ambient coordinates this is $x_2^2+x_3^2\le
x_1^2$) limited by the light-cone on which $\theta\to\infty$. The flat contraction $\k_1=0$ gives
rise to the usual spherical coordinates in the Euclidean space ($\k_2=1$).

By introducing the parametrization (\ref{da}) in the metric  written in terms of ambient
coordina\-tes~(\ref{ce}) we obtain that
\begin{gather}
{\rm d} s^2=
 {\rm d} r^2+   \k_2 \Sk_{\k_1}^2(r)\left( {\rm d} \theta^2 +\Sk_{\k_2}^2(\theta) {\rm d}
\phi^2 \right)  ,
\label{db}
\end{gather}
which is particularized in Table~1 to each space.
From it we compute the Levi-Civita connection~$\Gamma_{ij}^k$,    the Riemann $R^i_{jkl}$
and Ricci $R_{ij}$ tensors~\cite{Doub}. Their nonzero components are given~by
\begin{gather}
\Gamma^\theta_{\theta r}=\Gamma^\phi_{\phi r}=1/\Tk_{\k_1}(r),\qquad
\Gamma^\phi_{\phi\theta}=1/\Tk_{\k_2}(\theta),
\qquad \Gamma^r_{\theta\theta}=-\k_2\Sk_{\k_1}(r)\Ck_{\k_1}(r),\nonumber\\
\Gamma^r_{\phi\phi}=-\k_2\Sk_{\k_1}(r)\Ck_{\k_1}(r)\Sk_{\k_2}^2(\theta),\qquad
\Gamma^\theta_{\phi\phi}=-\Sk_{\k_2}(\theta)\Ck_{\k_2}(\theta),
\nonumber\\
R^r_{\theta r\theta}=R^\phi_{\theta \phi\theta}=\k_1\k_2\Sk_{\k_1}^2(r),\qquad
R^r_{\phi r\phi}=R^\theta_{\phi\theta \phi}=\k_1\k_2\Sk_{\k_1}^2(r)\Sk_{\k_2}^2(\theta),\qquad
R^\theta_{r\theta r}=R^\phi_{r\phi r}=\k_1,
\nonumber\\
R_{r r}=2\k_1,\qquad
R_{\theta\theta}=2\k_1\k_2\Sk_{\k_1}^2(r),\qquad
R_{\phi\phi}=2\k_1\k_2\Sk_{\k_1}^2(r)\Sk_{\k_2}^2(\theta).\label{conec}
\end{gather}
Therefore all the sectional curvatures
turn out to be   constant $K_{ij}=\k_1$, while the scalar curvature
reads  $K=6\k_1$.

\setcounter{equation}{0}

\section{Geodesic motion}

The metric (\ref{db}) can be read as the kinetic energy of a particle  written
in  terms of the velocities $(\dot r,\dot \theta,\dot \phi)$, that is, the Lagrangian of the
geodesic motion on the space ${\mathbb
S}^3_{[\k_1]\k_2}$ given by
\begin{gather}
{\cal T}=\frac 12\big(\dot r^2+ \k_2 \Sk_{\k_1}^2(r)\big( \dot\theta^2 +\Sk_{\k_2}^2(\theta)
\dot \phi^2 \big) \big).
\label{ea}
\end{gather}
Then the canonical momenta $(p_r,p_\theta,p_\phi)$ are obtained through $p=\partial {\cal
T}/\partial \dot q$ $(\dot q=\dot r,\dot \theta,\dot \phi)$, namely,
\begin{gather}
p_r=\dot r,\nonumber\\
p_\theta=\k_2 \Sk_{\k_1}^2(r) \dot\theta,\nonumber\\
p_\phi=\k_2 \Sk_{\k_1}^2(r) \Sk_{\k_2}^2(\theta) \dot \phi,
\label{eb}
\end{gather}
so that the free Hamiltonian in the geodesic polar phase space
$(q;p)=(r,\theta,\phi;p_r,p_\theta,p_\phi)$  with respect to the canonical Lie--Poisson bracket,
\begin{gather}
\left\{f,g \right\}
=\sum_{i=1}^3\left(\frac{\partial f}{\partial q_i}\frac{\partial g}{\partial p_i}
-\frac{\partial g}{\partial q_i}\frac{\partial f}{\partial p_i}\right),
\label{ec}
\end{gather}
turns out to be
\begin{gather}
{\cal T}=\frac 12\left(p_r^2+ \frac{p_\theta^2}{\k_2 \Sk_{\k_1}^2(r) }
+\frac{p_\phi^2 }{\k_2 \Sk_{\k_1}^2(r) \Sk_{\k_2}^2(\theta)}  \right).
\label{ed}
\end{gather}
Note that the connection (\ref{conec}) would allow one to write the geodesic equations whose
solution would correspond to the geodesic motion associated with ${\cal T}$ (see~\cite{jpa2D} for
the 2D case).

Now we proceed to deduce a phase space realization of the Lie generators of
$so_{\k_1,\k_2}(4)$. In Weierstrass coordinates
$x_\mu$ and momenta $p_\mu$ this comes from the vector fields (\ref{cf}) through the replacement
$\partial_\mu\to -p_\mu$:
\begin{alignat}{3}
&J_{01}=x_0 p_1-\k_1 x_1 p_0 ,\qquad &&
J_{0j}= x_0 p_j-\k_1\k_2 x_j p_0,&\nonumber\\
& J_{23}=  x_2 p_3-x_3 p_2 , \qquad  && J_{1j}= x_1 p_j-\k_2 x_j p_1. &
\label{eda}
\end{alignat}
The metric (\ref{ce}) can also be understood as the kinetic energy in the ambient velocities
$\dot x_\mu$ so that the   momenta $p_\mu$
are $(j=2,3)$:
\begin{gather}
p_0=\dot x_0/\k_1,\qquad p_1=\dot x_1,\qquad p_j=\k_2\dot x_j.
\label{edb}
\end{gather}
Next if we compute the velocities $\dot x_i$ in the parametrization (\ref{da}) and
introduce the momenta~(\ref{eb}) and~(\ref{edb})  we obtain   the relationship between the
ambient momenta and the geodesic polar ones:
\begin{gather*}
p_0=- \Sk_{\k_1}(r)\, p_r,\\
p_1=  \Ck_{\k_1}(r)
\Ck_{\k_2}(\theta)\, p_r-\frac{\Sk_{\k_2}(\theta)}{\Sk_{\k_1}(r)}\,p_\theta,\\
p_2= \k_2 \Ck_{\k_1}(r)
\Sk_{\k_2}(\theta)\cos\phi\,p_r
+\frac{\Ck_{\k_2}(\theta)\cos\phi}{\Sk_{\k_1}(r)}\,p_\theta -
\frac{\sin\phi}{\Sk_{\k_1}(r)\Sk_{\k_2}(\theta)}\,p_\phi ,\\
p_3= \k_2 \Ck_{\k_1}(r)
\Sk_{\k_2}(\theta)\sin\phi\,p_r
+\frac{\Ck_{\k_2}(\theta)\sin\phi}{\Sk_{\k_1}(r)}\,p_\theta +
\frac{\cos\phi}{\Sk_{\k_1}(r)\Sk_{\k_2}(\theta)}\,p_\phi .
\end{gather*}
 Hence the generators (\ref{eda})  in   geodesic polar
coordinates and momenta turn out to be
\begin{gather}
J_{01}=\Ck_{\k_2}(\theta)\, p_r-\frac{\Sk_{\k_2}(\theta)}{\Tk_{\k_1}(r)}\,
p_\theta,\nonumber\\
 J_{02}=\k_2\Sk_{\k_2}(\theta)\cos\phi\, p_r+
\frac{\Ck_{\k_2}(\theta)\cos\phi}{\Tk_{\k_1}(r)}\,
p_\theta-\frac{\sin\phi}{\Tk_{\k_1}(r)\Sk_{\k_2}(\theta)}\, p_\phi,\nonumber\\
 J_{03}=\k_2\Sk_{\k_2}(\theta)\sin\phi\, p_r+
\frac{\Ck_{\k_2}(\theta)\sin\phi}{\Tk_{\k_1}(r)}\,
p_\theta+\frac{\cos\phi}{\Tk_{\k_1}(r)\Sk_{\k_2}(\theta)}\, p_\phi,\nonumber\\
J_{12}=\cos\phi\, p_\theta-\frac{\sin\phi}{\Tk_{\k_2}(\theta)}\, p_\phi,\nonumber\\
J_{13}=\sin\phi\, p_\theta+\frac{\cos\phi}{\Tk_{\k_2}(\theta)}\, p_\phi,\nonumber\\
J_{23}=  p_\phi .
\label{ee}
\end{gather}

By direct computations it can be proven the following statement.

\begin{proposition}
The generators \eqref{ee} fulfil the commutation relations  \eqref{ba} with respect to the
Lie--Poisson bracket \eqref{ec} and all of them  Poisson commute with $\cal T$ \eqref{ed}.
\end{proposition}

In this respect, notice that, under~(\ref{ee}), the
kinetic energy is related with the Casimir~${\cal C}_1$~(\ref{bba})  by
$2\k_2{\cal T}={\cal C}_1$, while the second Casimir ${\cal C}_2$ vanishes.

The realization of the generators (\ref{ee}) is particularized for each specific space
and  Poisson--Lie algebra contained in ${\mathbb S}^3_{[\k_1]\k_2}$ and
$so_{\k_1,\k_2}(4)$ in Table~2. In order to present the simplest expressions,
hereafter we shall set in all the tables  $\k_1\in \{+1,0,-1\}$ and $\k_2\in\{+1,-1\}$, which
corresponds to deal with units $R=\tau=c=1$.

\bigskip

{\footnotesize

\noindent
{\bf Table 2.} Phase space realization of the generators of $so_{\k_1,\k_2}(4)$ in
canonical geodesic polar coordinates and momenta $(r,\theta,\phi;p_r,p_\theta,p_\phi)$ on each
space
 ${\mathbb S}^3_{[\k_1]\k_2}$   with $\k_1\in\{+1,0,-1\}$ and $\k_2\in\{+1,-1\}$.
$$
\begin{array}{ll}
\hline
\\[-6pt]
{\mbox {3D Riemannian spaces}}&\quad{\mbox  {$(2+1)$D
 Relativistic spacetimes}}\\[4pt]
\hline
\\[-6pt]
\mbox {$\bullet$ Spherical space  ${\mathbb S}^3_{[+]+}\equiv {\bf S}^3$:\quad
$so(4)$}&\quad\mbox {$\bullet$ Anti-de Sitter spacetime ${\mathbb S}^3_{[+]-}\equiv {\bf
AdS}^{2+1}$:\quad
$so(2,2)$}\\[5.5pt]
\displaystyle{J_{01}=\cos\theta\, p_r-\frac{\sin\theta}{\tan r}\,
p_\theta}&\quad \displaystyle{J_{01}=\cosh\theta\,
p_r-\frac{\sinh\theta}{\tan r}\, p_\theta}\\[5.5pt]
\displaystyle{ J_{02}= \sin \theta\cos\phi\, p_r+
\frac{\cos \theta\cos\phi}{\tan r}\,
p_\theta-\frac{\sin\phi\, p_\phi}{\tan r\sin \theta}}
&\quad \displaystyle{ J_{02}=-\sinh\theta\cos\phi\, p_r+
\frac{\cosh\theta\cos\phi}{\tan r}\,
p_\theta-\frac{\sin\phi\, p_\phi}{\tan r\sinh\theta}}\\[5.5pt]
\displaystyle{ J_{03}=\sin\theta\sin\phi\, p_r+
\frac{\cos\theta\sin\phi}{\tan r}\,
p_\theta+\frac{\cos\phi\, p_\phi}{\tan r\sin \theta}}&\quad \displaystyle{ J_{03}=-\sinh
\theta\sin\phi\, p_r+
\frac{\cosh\theta\sin\phi}{\tan r}\,
p_\theta+\frac{\cos\phi\, p_\phi}{\tan r\sinh\theta}}\\[5.5pt]
\displaystyle{J_{12}=\cos\phi\, p_\theta-\frac{\sin\phi}{\tan \theta}\,
p_\phi}&\quad \displaystyle{J_{12}=\cos\phi\, p_\theta-\frac{\sin\phi}{\tanh\theta}\,
p_\phi}\\[5.5pt]
\displaystyle{J_{13}=\sin\phi\, p_\theta+\frac{\cos\phi}{\tan\theta}\, p_\phi}&\quad
\displaystyle{J_{13}=\sin\phi\, p_\theta+\frac{\cos\phi}{\tanh\theta}\, p_\phi}\\[5.5pt]
\displaystyle{J_{23}=  p_\phi}&\quad \displaystyle{J_{23}=  p_\phi}\\[8pt]

\mbox {$\bullet$ Euclidean space  ${\mathbb S}^3_{[0]+}\equiv {\bf E}^3$:\quad
$iso(3)$}&\quad\mbox {$\bullet$ Minkowskian spacetime ${\mathbb S}^3_{[0]-}\equiv {\bf
M}^{2+1}$:\quad
$iso(2,1)$}\\[5.5pt]

\displaystyle{J_{01}=\cos\theta\, p_r-\frac{\sin\theta}{  r}\,
p_\theta}&\quad \displaystyle{J_{01}=\cosh\theta\,
p_r-\frac{\sinh\theta}{  r}\, p_\theta}\\[5.5pt]
\displaystyle{ J_{02}= \sin \theta\cos\phi\, p_r+
\frac{\cos \theta\cos\phi}{  r}\,
p_\theta-\frac{\sin\phi}{  r\sin \theta}\, p_\phi}
&\quad \displaystyle{ J_{02}=-\sinh\theta\cos\phi\, p_r+
\frac{\cosh\theta\cos\phi}{  r}\,
p_\theta-\frac{\sin\phi}{ r\sinh\theta}\, p_\phi}\\[5.5pt]
\displaystyle{ J_{03}=\sin\theta\sin\phi\, p_r+
\frac{\cos\theta\sin\phi}{  r}\,
p_\theta+\frac{\cos\phi}{ r\sin \theta}\, p_\phi}&\quad \displaystyle{ J_{03}=-\sinh
\theta\sin\phi\, p_r+
\frac{\cosh\theta\sin\phi}{  r}\,
p_\theta+\frac{\cos\phi}{  r\sinh\theta}\, p_\phi}\\[5.5pt]
\displaystyle{J_{12}=\cos\phi\, p_\theta-\frac{\sin\phi}{\tan \theta}\,
p_\phi}&\quad \displaystyle{J_{12}=\cos\phi\, p_\theta-\frac{\sin\phi}{\tanh\theta}\,
p_\phi}\\[5.5pt]
\displaystyle{J_{13}=\sin\phi\, p_\theta+\frac{\cos\phi}{\tan\theta}\, p_\phi}&\quad
\displaystyle{J_{13}=\sin\phi\, p_\theta+\frac{\cos\phi}{\tanh\theta}\, p_\phi}\\[5.5pt]
\displaystyle{J_{23}=  p_\phi}&\quad \displaystyle{J_{23}=  p_\phi}\\[8pt]

\mbox {$\bullet$  Hyperbolic space ${\mathbb S}^3_{[-]+}\equiv {\bf H}^3$:\quad
$so(3,1)$}&\quad\mbox {$\bullet$  De Sitter spacetime ${\mathbb S}^3_{[-]-}\equiv {\bf
dS}^{2+1}$:\quad $so(3,1)$}\\[4pt]

\displaystyle{J_{01}=\cos\theta\, p_r-\frac{\sin\theta}{\tanh r}\,
p_\theta}&\quad \displaystyle{J_{01}=\cosh\theta\,
p_r-\frac{\sinh\theta}{\tanh r}\, p_\theta}\\[5.5pt]
\displaystyle{ J_{02}= \sin \theta\cos\phi\, p_r+
\frac{\cos \theta\cos\phi}{\tanh r}\,
p_\theta-\frac{\sin\phi\, p_\phi}{\tanh r\sin \theta}}
&\quad \displaystyle{ J_{02}=-\sinh\theta\cos\phi\, p_r+
\frac{\cosh\theta\cos\phi}{\tanh r}\,
p_\theta-\frac{\sin\phi\, p_\phi}{\tanh r\sinh\theta}}\\[5.5pt]
\displaystyle{ J_{03}=\sin\theta\sin\phi\, p_r+
\frac{\cos\theta\sin\phi}{\tanh r}\,
p_\theta+\frac{\cos\phi\, p_\phi}{\tanh r\sin \theta}}&\quad \displaystyle{ J_{03}=-\sinh
\theta\sin\phi\, p_r+
\frac{\cosh\theta\sin\phi}{\tanh r}\,
p_\theta+\frac{\cos\phi\, p_\phi}{\tanh r\sinh\theta}}\\[5.5pt]
\displaystyle{J_{12}=\cos\phi\, p_\theta-\frac{\sin\phi}{\tan \theta}\,
p_\phi}&\quad \displaystyle{J_{12}=\cos\phi\, p_\theta-\frac{\sin\phi}{\tanh\theta}\,
p_\phi}\\[5.5pt]
\displaystyle{J_{13}=\sin\phi\, p_\theta+\frac{\cos\phi}{\tan\theta}\, p_\phi}&\quad
\displaystyle{J_{13}=\sin\phi\, p_\theta+\frac{\cos\phi}{\tanh\theta}\, p_\phi}\\[5.5pt]
\displaystyle{J_{23}=  p_\phi}&\quad \displaystyle{J_{23}=  p_\phi}\\[5pt]
\hline
\end{array}
$$
}

\vspace{-2mm}

\setcounter{equation}{0}

\section{Superintegrable potentials}

Now if we look for superintegrable potentials ${\cal U}(q)={\cal U}(r,\theta,\phi)$ which
generalize the Euclidean one (\ref{aa}) to the  space  ${\mathbb
S}^3_{[\k_1]\k_2}$ we find
\begin{gather}
{\cal U}={\cal
F}'(x_0)+\frac{\beta_1}{x_1^2}+\frac{\beta_2}{x_2^2}+\frac{\beta_3}{x_3^2}\nonumber\\
\phantom{{\cal U}}{} = {\cal F}(r)
+\frac{1}{\Sk_{\k_1}^2(r)}\left(\frac{\beta_1}{\Ck_{\k_2}^2(\theta)}+
\frac{\beta_2}{\Sk_{\k_2}^2(\theta)\cos^2\phi}
+\frac{\beta_3}{\Sk_{\k_2}^2(\theta)\sin^2\phi} \right),
\label{fa}
\end{gather}
where ${\cal F}'(\Ck_{\k_1}(r))\equiv {\cal
F}(r)$ is an arbitrary smooth function and $\beta_i$ are arbitrary real constants.
As in ${\bf E}^3$, the three $\beta_i$-terms can be interpreted on the six spaces in a common
way as ``centrifugal barriers''; for some particular curved spaces these may admit an
alternative interpretation as non-central harmonic oscillators. These facts will be
explained in detail in the next section.

\newpage

 The resulting Hamiltonian ${\cal H}={\cal T}+ {\cal U}$, with kinetic
energy (\ref{ed}) and potential (\ref{fa}), has three integrals of the motion
quadratic in the momenta which are associated with  the (Lorentz) rotation generators
 $(j=2,3)$:
\begin{gather}
 I_{1j}=J_{1j}^2+2\beta_1\k_2^2\frac{x_j^2}{x_1^2}+2\beta_j\k_2
\frac{x_1^2}{x_j^2} ,\qquad I_{23}=J_{23}^2+2\beta_2\k_2\frac{x_3^2}{x_2^2}+2\beta_3\k_2
\frac{x_2^2}{x_3^2} ,
\label{int}
\end{gather}
which in the geodesic polar phase space explicitly read
\begin{gather}
I_{12}=\left(\cos\phi\, p_\theta-\frac{\sin\phi}{\Tk_{\k_2}(\theta)}\,
p_\phi \right)^2+2\beta_1\k_2^2\Tk_{\k_2}^2(\theta)\cos^2\phi
+\frac{2\beta_2\k_2}{\Tk_{\k_2}^2(\theta)\cos^2\phi},\nonumber\\
I_{13}=\left(\sin\phi\, p_\theta+\frac{\cos\phi}{\Tk_{\k_2}(\theta)}\, p_\phi
\right)^2+2\beta_1\k_2^2\Tk_{\k_2}^2(\theta)\sin^2\phi
+\frac{2\beta_3\k_2}{\Tk_{\k_2}^2(\theta)\sin^2\phi},\nonumber\\
I_{23}=p_\phi^2+2\beta_2\k_2\tan^2\phi +\frac{2\beta_3\k_2}{\tan^2\phi} .
\label{fc}
\end{gather}
These constants of the motion do not Poisson commute with each
other. In order to find quantities in involution we define another
integral from the above set:
\begin{gather}
I_{123}=I_{12}+I_{13}+\k_2 I_{23}+2\k_2(\beta_1+\k_2\beta_2+\k_2\beta_3) \nonumber\\
\phantom{I_{123}}{}= p_\theta^2+\frac{p_\phi^2 }{
\Sk_{\k_2}^2(\theta)}
+\frac{2\beta_1\k_2}{\Ck_{\k_2}^2(\theta)}+\frac{2\beta_2\k_2}{\Sk_{\k_2}^2(\theta)\cos^2\phi}
+\frac{2\beta_3\k_2}{\Sk_{\k_2}^2(\theta)\sin^2\phi} ,
\label{fd}
\end{gather}
  which is
 related with the Casimir of the rotation subalgebra ${\mathfrak h}_0=so_{\k_2}(3)$.

Superintegrability of ${\cal H}$ is then characterized by:
\begin{proposition}
{\rm (i)} The three functions $\{I_{12},I_{123},{\cal H}\}$ are mutually in involution. The same
holds for the set $\{I_{23},I_{123},{\cal H}\}$.

{\rm (ii)} The four functions $\{I_{12},I_{23},I_{123},{\cal H}\}$ are functionally
independent, thus
${\cal H}$ is a  superintegrable Hamiltonian.
\end{proposition}

These results, which can be checked  directly,  are displayed in Table~3 for each
particular space arising within ${\mathbb S}^3_{[\k_1]\k_2}$. Notice that the integrals
$\{I_{12},I_{23},I_{123}\}$ do depend on $\k_2$ and $(\theta,\phi;p_\theta,p_\phi)$ but
neither on the curvature~$\k_1$ nor on
$(r,p_r)$, so these are the same for each set of three   spaces with the same signature.

A straightforward consequence of the complete integrability determined by
$\{I_{23},I_{123},{\cal H}\}$ is that ${\cal H}$ is separable and we   obtain  three
equations, each of them depending on a canonical pair
$(q_i,p_i)$:
\begin{gather}
I_{23}(\phi,p_\phi)=p_\phi^2+2\beta_2\k_2\tan^2\phi
+\frac{2\beta_3\k_2}{\tan^2\phi} ,\nonumber\\
I_{123}(\theta,p_\theta)=p_\theta^2
+\frac{2\beta_1\k_2}{\Ck_{\k_2}^2(\theta)}+\frac{1}{
\Sk_{\k_2}^2(\theta)}\left(I_{23}+2\k_2(\beta_2+\beta_3)
\right),\nonumber\\
{\cal H}(r,p_r)=\frac 12\, p_r^2+ {\cal F}(r)+ \frac{1}{2\k_2
\Sk_{\k_1}^2(r) } \,I_{123} ,
\label{fe}
\end{gather}
and ${\cal H}$ is so reduced to a 1D radial system.

Therefore there remains {\it one} constant of the motion to obtain
{\it maximal superintegrability} so that we shall say that  ${\cal H}$
is a {\it quasi-maximally superintegrable} Hamiltonian. In the next
sections we study two specific choices for the arbitrary radial function
${\cal F}(r)$ that lead to an additional integral thus providing
maximally superintegrable potentials. The resulting systems are
generalizations of the (curved) harmonic oscillator and KC
potentials with additional terms (dependending on the $\beta_i$).

\newpage

{\footnotesize
 \noindent
{\bf Table 3.} Superintegrable Hamiltonian ${\cal H}={\cal T}+{\cal U}$ and its three constants of
the motion $\{I_{12},I_{23},I_{123}\}$ for the six spaces ${\mathbb
S}^3_{[\k_1]\k_2}$ with $\k_1\in\{+1,0,-1\}$ and $\k_2\in\{+1,-1\}$.
$$
\begin{array}{l}
\hline
\\[-7pt]
\multicolumn{1}{c}{\mbox {3D Riemannian spaces}} \\[3pt]
\hline
\\[-7pt]
\mbox {$\bullet$ Spherical space ${\mathbb S}^3_{[+]+}\equiv {\bf S}^3$} \\[4pt]
\displaystyle{{\cal H}=\frac 12\left(p_r^2+ \frac{p_\theta^2}{  \sin^2 r
}  +\frac{p_\phi^2 }{  \sin^2 r \sin^2\theta}  \right)
+ {\cal F}(r)
+\frac{1}{\sin^2 r}\left(\frac{\beta_1}{\cos^2\theta}+
\frac{\beta_2}{\sin^2\theta\cos^2\phi}
+\frac{\beta_3}{\sin^2\theta\sin^2\phi} \right)}\\[10pt]
\mbox {$\bullet$ Euclidean space  ${\mathbb S}^3_{[0]+}\equiv {\bf E}^3$ } \\[4pt]
\displaystyle{{\cal H}=\frac 12\left(p_r^2+ \frac{p_\theta^2}{ r^2
}  +\frac{p_\phi^2 }{  r^2 \sin^2\theta}  \right)
+ {\cal F}(r)
+\frac{1}{r^2}\left(\frac{\beta_1}{\cos^2\theta}+
\frac{\beta_2}{\sin^2\theta\cos^2\phi}
+\frac{\beta_3}{\sin^2\theta\sin^2\phi} \right)}\\[10pt]
\mbox {$\bullet$ Hyperbolic space ${\mathbb S}^3_{[-]+}\equiv{\bf H}^3$  }\\[4pt]
\displaystyle{{\cal H}=\frac 12\left(p_r^2+ \frac{p_\theta^2}{  \sinh^2 r
}  +\frac{p_\phi^2 }{  \sinh^2 r \sin^2\theta}  \right)
+ {\cal F}(r)
+\frac{1}{\sinh^2 r}\left(\frac{\beta_1}{\cos^2\theta}+
\frac{\beta_2}{\sin^2\theta\cos^2\phi}
+\frac{\beta_3}{\sin^2\theta\sin^2\phi} \right)}\\[16pt]
\displaystyle{I_{12}=\left(\cos\phi\, p_\theta-\frac{\sin\phi}{\tan\theta}\,
p_\phi \right)^2+2\beta_1 \tan^2\theta\cos^2\phi
+\frac{2\beta_2 }{\tan^2\theta\cos^2\phi}}\\[8pt]
\displaystyle{I_{23}=p_\phi^2+2\beta_2 \tan^2\phi
+\frac{2\beta_3 }{\tan^2\phi} }\\
\displaystyle{I_{123} = p_\theta^2+\frac{p_\phi^2 }{
\sin^2\theta}
+\frac{2\beta_1 }{\cos^2\theta}
+\frac{2\beta_2 }{\sin^2\theta\cos^2\phi}
+\frac{2\beta_3 }{\sin^2\theta\sin^2\phi} }\\[10pt]
 \hline
\\[-7pt]
\multicolumn{1}{c}{\mbox {$(2+1)$D
 Relativistic spacetimes
}} \\[3pt]
\hline
\\[-7pt]
\mbox {$\bullet$ Anti-de Sitter
spacetime ${\mathbb S}^3_{[+]-}\equiv {\bf AdS}^{2+1}$   } \\[4pt]
\displaystyle{{\cal H}=\frac 12\left(p_r^2- \frac{p_\theta^2}{  \sin^2 r}
-\frac{p_\phi^2 }{ \sin^2 r \sinh^2\theta}  \right)
+ {\cal F}(r)
+\frac{1}{\sin^2 r}\left(\frac{\beta_1}{\cosh^2\theta}+
\frac{\beta_2}{\sinh^2\theta\cos^2\phi}
+\frac{\beta_3}{\sinh^2\theta\sin^2\phi} \right)}\\[10pt]
\mbox {$\bullet$ Minkowskian spacetime ${\mathbb S}^3_{[0]-}\equiv{\bf
M}^{2+1}$  } \\[4pt]
\displaystyle{{\cal H}=\frac 12\left(p_r^2- \frac{p_\theta^2}{ r^2}
-\frac{p_\phi^2 }{r^2 \sinh^2\theta}  \right)
+ {\cal F}(r)
+\frac{1}{r^2}\left(\frac{\beta_1}{\cosh^2\theta}+
\frac{\beta_2}{\sinh^2\theta\cos^2\phi}
+\frac{\beta_3}{\sinh^2\theta\sin^2\phi} \right)}\\[10pt]
\mbox {$\bullet$ De Sitter
spacetime
 ${\mathbb S}^3_{[-]-}\equiv{\bf dS}^{2+1}$  }\\[4pt]
\displaystyle{{\cal H}=\frac 12\left(p_r^2- \frac{p_\theta^2}{  \sinh^2 r }
-\frac{p_\phi^2 }{ \sinh^2 r \sinh^2\theta}  \right)
+ {\cal F}(r)
+\frac{1}{\sinh^2 r}\left(\frac{\beta_1}{\cosh^2\theta}+
\frac{\beta_2}{\sinh^2\theta\cos^2\phi}
+\frac{\beta_3}{\sinh^2\theta\sin^2\phi} \right)}\\[16pt]
\displaystyle{I_{12}=\left(\cos\phi\, p_\theta-\frac{\sin\phi}{\tanh\theta}\,
p_\phi \right)^2+2\beta_1 \tanh^2\theta\cos^2\phi
-\frac{2\beta_2 }{\tanh^2\theta\cos^2\phi}}\\[8pt]
\displaystyle{I_{23}=p_\phi^2-2\beta_2 \tan^2\phi
-\frac{2\beta_3}{\tan^2\phi} }\\
\displaystyle{I_{123} = p_\theta^2+\frac{p_\phi^2 }{
\sinh^2\theta}
-\frac{2\beta_1 }{\cosh^2\theta}
-\frac{2\beta_2 }{\sinh^2\theta\cos^2\phi}
-\frac{2\beta_3 }{\sinh^2\theta\sin^2\phi} }\\[10pt]
\hline
\end{array}
$$
}

\setcounter{equation}{0}

\section{Harmonic oscillator potential}

If we like   to extend the (curved) harmonic oscillator potential (\ref{ad}) to our six
spaces, we have to consider the following   choice for the arbitrary function appearing in
(\ref{fa}):
\begin{gather}
{\cal
F}'(x_0)=\beta_0\left(\frac{1-x_0^2}{\k_1 x_0^2} \right)=
\beta_0\left(\frac{ x_1^2+ \k_2 x_2^2 + \k_2 x_3^2 }{x_0^2} \right),\qquad
{\cal F}(r)=\beta_0\Tk^2_{\k_1}(r),
\label{ga}
\end{gather}
where $\beta_0$ is an arbitrary real parameter. When the complete Hamiltonian is considered
  we obtain the generalization of 3D  SW system~(\ref{ab}), ${\cal H}^{\rm
SW}={\cal T}+{\cal U}^{\rm SW}$, to the space
${\mathbb S}^3_{[\k_1]\k_2}$, namely
\begin{gather}
{\cal U}^{\rm
SW}=\beta_0\Tk^2_{\k_1}(r)+\frac{1}{\Sk_{\k_1}^2(r)}\left(\frac{\beta_1}{\Ck_{\k_2}^2(\theta)}+
\frac{\beta_2}{\Sk_{\k_2}^2(\theta)\cos^2\phi}
+\frac{\beta_3}{\Sk_{\k_2}^2(\theta)\sin^2\phi} \right).
\label{gb}
\end{gather}
As we already mentioned in the introduction,   the proper SW Hamiltonian arises in the (flat)
Euclidean space~\cite{7,8,9,10}, here written in polar coordinates, which is formed by an
isotropic harmonic oscillator with angular frequency $\omega=\sqrt{\beta_0}$ together with
three centrifugal barriers associated with the $\beta_i$'s. Different constructions of the
SW system on the (curved) spherical and hyperbolic spaces can be found in~\cite{VulpiLett,
11,CRMVulpi,18,20,21,27}. More recently such a potential has also been deduced and
analysed in the $(1+1)$D relativistic spacetimes of constant curvature in~\cite{jpa2D,car2} as
well as    in 2D spaces of variable curvature in~\cite{jpa2D}.

In our case, any of the translation generators $J_{0i}$ (\ref{ee})  provides a constant of
the motion quadratic in the momenta in the form
 $(j=2,3)$:
\begin{gather}
I_{01}=J_{01}^2+2\beta_0\frac{x_1^2}{x_0^2}+2\beta_1
\frac{x_1^2}{x_0^2} ,\nonumber\\
 I_{0j}=J_{0j}^2+2\beta_0\k_2^2\frac{x_j^2}{x_0^2}+2\beta_j\k_2
\frac{x_0^2}{x_j^2} ,
\label{inta}
\end{gather}
that is,
\begin{gather}
I_{01}=J_{01}^2+2\beta_0 \Tk_{\k_1}^2(r)\Ck_{\k_2}^2(\theta)+
 \frac{2\beta_1}{\Tk_{\k_1}^2(r)\Ck_{\k_2}^2(\theta)} ,\nonumber\\
I_{02}=J_{02}^2+2\beta_0\k_2^2
\Tk_{\k_1}^2(r)\Sk_{\k_2}^2(\theta)\cos^2\phi+
\frac{2\beta_2\k_2}{\Tk_{\k_1}^2(r)\Sk_{\k_2}^2(\theta)\cos^2\phi} ,\nonumber\\
I_{03}=J_{03}^2+2\beta_0\k_2^2
\Tk_{\k_1}^2(r)\Sk_{\k_2}^2(\theta)\sin^2\phi+
\frac{2\beta_3\k_2}{\Tk_{\k_1}^2(r)\Sk_{\k_2}^2(\theta)\sin^2\phi} .
\label{gd}
\end{gather}

Obviously, the seven integrals of the motion $\{I_{01},I_{02},I_{03},I_{12},I_{23},I_{123},
{\cal H}^{\rm SW}\}$ cannot be  functionally independent. One constraint for them is
given by
\begin{gather*}
2\k_2{\cal H}^{\rm SW}=\k_2 I_{01}+I_{02}+I_{03}+\k_1 I_{123},
\end{gather*}
which reminds the aforementioned relation for the geodesic motion $2\k_2{\cal T}={\cal
C}_1$. Note also that
\begin{gather*}
\{I_{01},I_{23}\}=\{I_{02},I_{13}\}=\{I_{03},I_{12}\}=0.
\end{gather*}

 The final result concerning the superintegrability of ${\cal H}^{\rm SW}$ is
established by:
\begin{proposition}
{\rm (i)} Each function $I_{0i}$ \eqref{gd} $(i=1,2,3)$ Poisson commutes with   ${\cal
H}^{\rm SW}$.

 {\rm (ii)} The five  functions $\{I_{0i},I_{12},I_{23},I_{123},{\cal H}^{\rm SW}\}$, where
$i$ is fixed, are functionally independent, thus
${\cal H}^{\rm SW}$ is a  maximally superintegrable Hamiltonian.
\end{proposition}

 The  Hamiltonian ${\cal H}^{\rm SW}$
and the additional constant of the motion $I_{01}$ (that ensures   maximal
superintegrability) are presented for each particular space contained in ${\mathbb
S}^3_{[\k_1]\k_2}$ in Table~4.

\subsection{Description of the  SW potential}

The 2D version of the potential  ${\cal U}^{\rm SW}$ (\ref{gb}) on ${\bf S}^2$ has been
interpreted in~\cite{ran,ran1,ran2} as a~superposition of three spherical oscillators; the
interpretation for arbitrary dimension on  ${\bf S}^N$ and ${\bf H}^N$ has been presented
in~\cite{VulpiLett,CRMVulpi}. Furthermore, a detail description on this potential on
 ${\bf AdS}^{1+1}$, ${\bf M}^{1+1}$ and ${\bf dS}^{1+1}$ was recently performed
in~\cite{jpa2D}.  In what follows we analyse the
(physical) geo\-met\-rical role of the 3D potential (\ref{gb}) on each particular space ${\mathbb
S}^3_{[\k_1]\k_2}$ thus generalizing    all of the mentioned 2D results.

\bigskip

{\footnotesize
 \noindent
{\bf Table 4.} Maximally superintegrable Smorodinsky--Winternitz Hamiltonian ${\cal H}^{\rm
SW}={\cal T}+{\cal U}^{\rm
SW}$ and the additional
  constant of the motion $I_{01}$ to the set $\{I_{12},I_{23},I_{123}\}$ for the six spaces
${\mathbb S}^3_{[\k_1]\k_2}$ with the same conventions given in Table~3.
$$
\begin{array}{l}
\hline
\\[-7pt]
\multicolumn{1}{c}{\mbox {3D Riemannian spaces}} \\[3pt]
\hline
\\[-7pt]
\mbox {$\bullet$ Spherical space ${\bf S}^3$} \\[4pt]
\displaystyle{{\cal H}^{\rm SW}=\frac 12\left(p_r^2+ \frac{p_\theta^2}{  \sin^2 r
}  +\frac{p_\phi^2 }{  \sin^2 r \sin^2\theta}  \right)
+ \beta_0\tan^2 r
+\frac{1}{\sin^2 r}\left(\frac{\beta_1}{\cos^2\theta}+
\frac{\beta_2}{\sin^2\theta\cos^2\phi}
+\frac{\beta_3}{\sin^2\theta\sin^2\phi} \right)}\\[9pt]
\displaystyle{ I_{01}=\left( \cos\theta\,
p_r-\frac{\sin\theta}{\tan r}\, p_\theta\right)^2+2\beta_0
\tan^2 r\cos^2\theta+
 \frac{2\beta_1}{\tan^2 r\cos^2\theta}   }\\[12pt]
\mbox {$\bullet$ Euclidean space  ${\bf E}^3$ } \\[4pt]
\displaystyle{{\cal H}^{\rm SW}=\frac 12\left(p_r^2+ \frac{p_\theta^2}{ r^2
}  +\frac{p_\phi^2 }{  r^2 \sin^2\theta}  \right)
+ \beta_0\, r^2
+\frac{1}{r^2}\left(\frac{\beta_1}{\cos^2\theta}+
\frac{\beta_2}{\sin^2\theta\cos^2\phi}
+\frac{\beta_3}{\sin^2\theta\sin^2\phi} \right)}\\[9pt]
\displaystyle{ I_{01}=\left( \cos\theta\,
p_r-\frac{\sin\theta}{  r}\, p_\theta\right)^2+2\beta_0\,
  r^2\cos^2\theta+
 \frac{2\beta_1}{r^2\cos^2\theta}   }\\[12pt]
\mbox {$\bullet$ Hyperbolic space ${\bf H}^3$  }\\[4pt]
\displaystyle{{\cal H}^{\rm SW}=\frac 12\left(p_r^2+ \frac{p_\theta^2}{  \sinh^2 r
}  +\frac{p_\phi^2 }{  \sinh^2 r \sin^2\theta}  \right)
+ \beta_0\tanh^2 r
+\frac{1}{\sinh^2 r}\left(\frac{\beta_1}{\cos^2\theta}+
\frac{\beta_2}{\sin^2\theta\cos^2\phi}
+\frac{\beta_3}{\sin^2\theta\sin^2\phi} \right)}\\[9pt]
\displaystyle{ I_{01}=\left( \cos\theta\,
p_r-\frac{\sin\theta}{\tanh r}\, p_\theta\right)^2+2\beta_0
\tanh^2 r\cos^2\theta+
 \frac{2\beta_1}{\tanh^2 r\cos^2\theta}   }\\[9pt]
 \hline
\\[-7pt]
\multicolumn{1}{c}{\mbox {$(2+1)$D
 Relativistic spacetimes
}} \\[3pt]
\hline
\\[-7pt]
\mbox {$\bullet$ Anti-de Sitter
spacetime $ {\bf AdS}^{2+1}$   } \\[4pt]
\displaystyle{{\cal H}^{\rm SW}=\frac 12\left(p_r^2- \frac{p_\theta^2}{  \sin^2 r}
-\frac{p_\phi^2 }{ \sin^2 r \sinh^2\theta}  \right)
+ \beta_0\tan^2 r
+\frac{1}{\sin^2 r}\left(\frac{\beta_1}{\cosh^2\theta}+
\frac{\beta_2}{\sinh^2\theta\cos^2\phi}
+\frac{\beta_3}{\sinh^2\theta\sin^2\phi} \right)}\\[9pt]
\displaystyle{ I_{01}=\left( \cosh\theta\,
p_r-\frac{\sinh\theta}{\tan r}\, p_\theta\right)^2+2\beta_0
\tan^2 r\cosh^2\theta+
 \frac{2\beta_1}{\tan^2 r\cosh^2\theta}   }\\[12pt]
\mbox {$\bullet$ Minkowskian spacetime ${\bf
M}^{2+1}$  } \\[4pt]
\displaystyle{{\cal H}^{\rm SW}=\frac 12\left(p_r^2- \frac{p_\theta^2}{ r^2}
-\frac{p_\phi^2 }{r^2 \sinh^2\theta}  \right)
+ \beta_0\, r^2
+\frac{1}{r^2}\left(\frac{\beta_1}{\cosh^2\theta}+
\frac{\beta_2}{\sinh^2\theta\cos^2\phi}
+\frac{\beta_3}{\sinh^2\theta\sin^2\phi} \right)}\\[9pt]
\displaystyle{ I_{01}=\left( \cosh\theta\,
p_r-\frac{\sinh\theta}{  r}\, p_\theta\right)^2+2\beta_0\,
  r^2\cosh^2\theta+
 \frac{2\beta_1}{r^2\cosh^2\theta}   }\\[12pt]
\mbox {$\bullet$ De Sitter
spacetime
 $ {\bf dS}^{2+1}$  }\\[4pt]
\displaystyle{{\cal H}^{\rm SW}=\frac 12\left(p_r^2- \frac{p_\theta^2}{  \sinh^2 r }
-\frac{p_\phi^2 }{ \sinh^2 r \sinh^2\theta}  \right)
+ \beta_0\tanh^2 r
+\frac{1}{\sinh^2 r}\left(\frac{\beta_1}{\cosh^2\theta}+
\frac{\beta_2}{\sinh^2\theta\cos^2\phi}
+\frac{\beta_3}{\sinh^2\theta\sin^2\phi} \right)}\\[9pt]
\displaystyle{ I_{01}=\left( \cosh\theta\,
p_r-\frac{\sinh\theta}{\tanh r}\, p_\theta\right)^2+2\beta_0
\tanh^2 r\cosh^2\theta+
 \frac{2\beta_1}{\tanh^2 r\cosh^2\theta}   }\\[9pt]
\hline
\end{array}
$$
}

Consider the   (time-like) geodesic $l_1$ and the two   (space-like)
geodesics $l_2$, $l_3$  in ${\mathbb S}^3_{[\k_1]\k_2}$ orthogonal at the origin $O$ and the
generic point
$Q(r,\theta,\phi)$ as given in Subsection~2.2. Next let $Q_{ij}$ $(i,j=1,2,3;\ i<j)$ be the
intersection point of the reference flag spanned by $l_i$ and $l_j$ (the 2-plane $l_il_j$)
with   its  orthogonal geodesic    through $Q$. Hence we introduce the (time-like) geodesic
distance
$x=Q Q_{23}$ and the two (space-like)   distances $y=Q Q_{13}$, $z=Q Q_{12}$. Finally, let
$Q_1$ be the intersection point of $l_1$ with its orthogonal (space-like) geodesic $l'_1$
through $Q$ for which $h= Q Q_1$ is the (space-like) distance measured along $l'_1$.
Now by applying trigonometry~\cite{trigo} on the orthogonal triangles $OQQ_1$ (with inner
angle $\theta$), $OQQ_{23}$ (with external
angle $\theta$), $Q_{13}QQ_1$ (with external
angle $\phi$) and $Q_{12}QQ_1$ (with inner
angle $\phi$),  we find that
\begin{alignat*}{3}
&OQQ_1:\quad&& \Sk_{\k_1\k_2}(h)=\Sk_{\k_1}(r)\Sk_{\k_2}(\theta),&\\
&OQQ_{23}:\quad && \Sk_{\k_1}(x)=\Sk_{\k_1}(r)\Ck_{\k_2}(\theta),&\\
& Q_{13}QQ_1:&& \Sk_{\k_1\k_2}(y)=\Sk_{\k_1\k_2}(h)\cos\phi,&\\
& Q_{12}QQ_1:\quad && \Sk_{\k_1\k_2}(z)=\Sk_{\k_1\k_2}(h)\sin\phi.&
\end{alignat*}
Hence the ambient coordinates $x_i$ (\ref{da}) can be expressed as
\begin{gather}
x_1=\Sk_{\k_1}(r)\Ck_{\k_2}(\theta)=\Sk_{\k_1}(x)  ,\nonumber\\
x_2= \Sk_{\k_1}(r)\Sk_{\k_2}(\theta)\cos\phi= \Sk_{\k_1\k_2}(y),\nonumber\\
x_3=\Sk_{\k_1}(r)\Sk_{\k_2}(\theta)\sin\phi= \Sk_{\k_1\k_2}(z),
\label{ggff}
\end{gather}
so that the SW potential (\ref{gb}) can be rewritten as
\begin{gather}
{\cal U}^{\rm
SW}=\beta_0\Tk^2_{\k_1}(r)+ \frac{\beta_1}{\Sk_{\k_1}^2 (x) }+
\frac{\beta_2}{\Sk_{\k_1\k_2}^2(y)}
+\frac{\beta_3}{\Sk_{\k_1\k_2}^2(z) }  ,
\label{gf}
\end{gather}
 which allows for a unified interpretation on the six spaces:
\begin{itemize}
\itemsep=0pt
\item  The $\beta_0$-term   is a {\it central} harmonic oscillator, that is, the
Higgs oscillator~\cite{Higgs} with center at the origin $O$.
\item  The three $\beta_i$-terms $(i=1,2,3)$ are ``centrifugal  barriers".
 \end{itemize}

Furthermore, the   $\beta_i$-potentials can be interpreted as non-central oscillators in some
particular spaces that we proceed to describe by considering the simplest values for
$\k_i\in\{\pm 1\}$.


\subsubsection[Spherical space  ${\bf S}^3$]{Spherical space  $\boldsymbol{{\bf S}^3}$}

  Let
$O_i$ be the points placed along the basic geodesics $l_i$ $(i=1,2,3)$ which are a quadrant
apart from the origin $O$, that is, each two points taken from the set $\{O,O_i\}$ are
mutually separated a~distance $\frac{\pi}2$ (if $\k_1=1/R^2$, a quadrant is
$\pi/(2\sqrt{\k_1})=R\pi/2$). In fact, each   $O_i$ is the intersection point   between
the   geodesic $l_i$ and the  axis $x_i$ of the ambient space. If we  denote by
$r_i$ the~distance between $Q$ and $O_i$ measured along the geodesic joining both points then
\[
r_1+x=r_2+y=r_3+z=\frac{\pi}2 ,
\]
 which means that each set of three points $\{O_1QQ_{23} \}$, $\{O_2QQ_{13} \}$ and
$\{O_3QQ_{12} \}$ lie on the same geodesic. Thus
\[
x_1=\sin x =\cos r_1,\qquad
x_2=\sin y =\cos r_2,\qquad
x_3=\sin z =\cos r_3,
\]
so that the SW potential  (\ref{gb})  on ${\bf S}^3$ can be expressed in two manners
\begin{gather}
{\cal U}^{\rm SW}=\beta_0\tan^2 r+\frac{\beta_1}{\sin^2
x}+\frac{\beta_2}{\sin^2 y}+\frac{\beta_3}{\sin^2 z}\label{gga}\\
\phantom{{\cal U}^{\rm SW}}{}  =   \beta_0\tan^2 r+\sum_{i=1}^3\left( \beta_i\tan^2 r_i+\beta_i \right),
\label{ggb}
\end{gather}
which show a superposition of the central spherical oscillator with center at $O$   either
with three spherical centrifugal barriers, or with three  spherical oscillators with centers
 placed at~$O_i$~\cite{VulpiLett,CRMVulpi}.

\subsubsection[Hyperbolic space  ${\bf H}^3$]{Hyperbolic space  $\boldsymbol{{\bf H}^3}$}

The analogous points to the previous ``centers"
$O_i$   would be beyond the ``actual" hyperbolic space and so
placed  in the exterior (``ideal") region of ${\bf H}^3$. The
SW potential can only written in the form~(\ref{gf}):
\begin{gather}
{\cal U}^{\rm
SW}=\beta_0\tanh^2 r+\frac{\beta_1}{\sinh^2
x}+\frac{\beta_2}{\sinh^2 y} +\frac{\beta_3}{\sinh^2 z} ,
\label{gh}
\end{gather}
giving rise to the superposition of a central hyperbolic oscillator
with three hyperbolic centrifugal barriers~\cite{CRMVulpi}.


\subsubsection[Euclidean space  ${\bf E}^3$]{Euclidean space  $\boldsymbol{{\bf E}^3}$}

The contraction $\k_1\to 0$ $(R\to \infty)$ of the SW  potential   on   ${\bf
S}^3$ and ${\bf H}^3$  can  be applied on both expressions (\ref{gga}) and (\ref{gh})
reducing to
\begin{gather}
{\cal U}^{\rm SW}=\beta_0  r^2+\frac{\beta_1}{x^2}+\frac{\beta_2}{y^2} +\frac{\beta_3}{z^2} ,
\label{gi}
\end{gather}
which is just the proper SW potential (\ref{ab}) formed by the flat  harmonic oscillator
with three centrifugal barriers; in this case
$(x,y,z)$ are Cartesian coordinates on ${\bf E}^3$ and $r^2=x^2+y^2+z^2$.
This contraction cannot   be performed on ${\bf S}^3$ when the   potential   is written
in the form (\ref{ggb}); notice that if $\k_1\to 0$ the points $O_i\to
\infty$.

\subsubsection[Anti-de Sitter spacetime ${\bf AdS}^{2+1}$]{Anti-de Sitter spacetime $\boldsymbol{{\bf AdS}^{2+1}}$}

We consider the intersection point $O_1$ between the time-like geodesic
$l_1$ and the  axis $x_1$ of the ambient space,  which is at a time-like distance $\frac
\pi 2$ from the origin
$O$~\cite{jpa2D}.  If $r_1$  denotes  the
time-like distance  $QO_1$, then $r_1+x=\frac \pi 2$.   Therefore the  SW potential becomes
\begin{gather}
{\cal U}^{\rm SW}=\beta_0\tan^2 r+\frac{\beta_1}{\sin^2
x}+\frac{\beta_2}{\sinh^2 y}+\frac{\beta_3}{\sinh^2 z}\label{gka}\\
\phantom{{\cal U}^{\rm SW}}{}  =  \beta_0\tan^2 r+\beta_1\tan^2 r_1+\frac{\beta_2}{\sinh^2
y}+\frac{\beta_3}{\sinh^2 z}+\beta_1 .
\label{gkb}
\end{gather}
The former expression corresponds to  the superposition of  a time-like (spherical)
oscillator centered at $O$  with a time-like (spherical) centrifugal potential and two
space-like (hyperbolic) ones. Under the latter  form, the
time-like centrifugal term is transformed into another
  spherical oscillator now with center at $O_1$.

\subsubsection[De Sitter spacetime ${\bf dS}^{2+1}$]{De Sitter spacetime $\boldsymbol{{\bf dS}^{2+1}}$}

Recall that ${\bf AdS}^{2+1}$ and  ${\bf dS}^{2+1}$ are related through an
  interchange between time-like lines and space-like ones; the former are compact
(circular) on
${\bf AdS}^{2+1}$ and non-compact (hyperbolic) on ${\bf dS}^{2+1}$, while the latter
are  non-compact on   ${\bf AdS}^{2+1}$ but  compact on ${\bf dS}^{2+1}$.

So,   we consider the intersection point $O_j$ $(j=2,3)$ between the space-like geodesic
$l_j$ and the  axis $x_j$ which is at a space-like distance $\frac \pi 2$ from   $O$,
so that  $r_j$ is  the space-like distance  $QO_j$ verifying
$r_2+y=r_3+z=\frac{\pi}2$~\cite{jpa2D}. Hence  the   SW  potential can be rewritten as
\begin{gather}
{\cal U}^{\rm SW}=\beta_0\tanh^2 r+\frac{\beta_1}{\sinh^2
x}+\frac{\beta_2}{\sin^2 y}+\frac{\beta_3}{\sin^2 z}\label{gna}\\
\phantom{{\cal U}^{\rm SW}}{}  =  \beta_0\tanh^2 r+\frac{\beta_1}{\sinh^2
x}+\beta_2\tan^2 r_2+\beta_3\tan^2 r_3+\beta_2+\beta_3 .
\label{gnb}
\end{gather}
In this way, we find the superposition of a central time-like (hyperbolic) oscillator with
a time-like (hyperbolic) centrifugal barrier, and  either with   two other space-like
(spherical) centrifugal barriers or with two   space-like (spherical)
oscillators   centered at $O_j$.

\subsubsection[Minkowskian spacetime ${\bf M}^{2+1}$]{Minkowskian spacetime $\boldsymbol{{\bf M}^{2+1}}$}

Finally, the
contraction $\k_1\to 0$ $(\tau\to \infty)$ of (\ref{gka}) and (\ref{gna})  gives
\begin{gather}
{\cal U}^{\rm SW}=\beta_0  r^2+\frac{\beta_1}{x^2}+\frac{\beta_2}{y^2} +\frac{\beta_3}{z^2} ,
\label{gij}
\end{gather}
  which is formed by a time-like
harmonic oscillator $\beta_0 r^2$, one time-like  centrifugal
barrier $\beta_1  / x^2$ together with two  space-like ones    $\beta_2  / y^2$, $\beta_3/
z^2$. The coordinates $(x,y,z)$ are the usual time and space ones such that
$ r^2= x^2-y^2-z^2$. On the
contrary, the expressions (\ref{gkb}) and (\ref{gnb}) are not well defined when
$\k_1\to 0$ since   the points $O_1$ and $O_j$ go to infinity.

\setcounter{equation}{0}

\section[Kepler-Coulomb potential]{Kepler--Coulomb potential}

The generalization of the KC potential (\ref{ad}) to the space ${\mathbb
S}^3_{[\k_1]\k_2}$ is achieved by choosing
\begin{gather}
{\cal
F}'(x_0)=-k \,\frac{x_0}{\sqrt{(1-x_0^2)/\k_1}} =
-k\, \frac{x_0}{\sqrt{x_1^2+\k_2 x_2^2 +\k_2 x_3^2}},\qquad
{\cal F}(r)=-\frac{k }{\Tk_{\k_1}(r)},
\label{ha}
\end{gather}
where $k$ is an arbitrary real parameter. As it already happens in
${\bf E}^3$~\cite{6}, it is not possible to add the three
potential terms depending on the $\beta_i$'s keeping at the same
time maximal superintegrability; so that, at least, one of them
must vanish. Consequently, we find, in principle, three possible
generalizations of the Euclidean potential (\ref{ac}) to ${\mathbb
S}^3_{[\k_1]\k_2}$:
\begin{gather}
{\cal U}^{\rm
GKC}_1=-\frac{k
}{\Tk_{\k_1}(r)}+\frac{1}{\Sk_{\k_1}^2(r)\Sk_{\k_2}^2(\theta)}\left(
\frac{\beta_2}{\cos^2\phi}
+\frac{\beta_3}{ \sin^2\phi} \right),\nonumber\\
{\cal U}^{\rm GKC}_2=-\frac{k
}{\Tk_{\k_1}(r)}+\frac{1}{\Sk_{\k_1}^2(r)}\left(\frac{\beta_1}{\Ck_{\k_2}^2(\theta)}+
\frac{\beta_3}{\Sk_{\k_2}^2(\theta)\sin^2\phi} \right),\nonumber\\
{\cal U}^{\rm GKC}_3=-\frac{k
}{\Tk_{\k_1}(r)}+\frac{1}{\Sk_{\k_1}^2(r)}\left(\frac{\beta_1}{\Ck_{\k_2}^2(\theta)}+
\frac{\beta_2}{\Sk_{\k_2}^2(\theta)\cos^2\phi} \right).
\label{hb}
\end{gather}
Thus each potential ${\cal U}^{\rm GKC}_i$ contains  the proper KC
$k$-term~\cite{car1,car2,18,21,
Schrodingerdual,Schrodingerdualc,Schrodingerdualb,27,Schrodinger} together with two
additional $\beta_i$-terms, which can further be  interpreted as centrifugal barriers or
non-central oscillators; for each of them   there is an additional constant of the motion
given by $(i=1,2,3)$:
\begin{gather}
 L_i=\sum_{l=1; l\ne i}^3J_{0l}J_{li}+k\,\frac{\k_2
x_i}{\sqrt{x_1^2+\k_2 x_2^2 +\k_2
x_3^2}}-2\k_2\sum_{l=1; l\ne i}^3 \beta_l\frac{x_0x_i}{x_l^2} ,
\label{intb}
\end{gather}
where $J_{li}=-J_{il}$ for $i<l$. In terms of the geodesic polar phase space these
integrals read
 \begin{gather}
L_1=-J_{02}J_{12}-J_{03}J_{13}+k\, \k_2\Ck_{\k_2}(\theta)
-\frac{2\k_2\Ck_{\k_2}(\theta)}{\Tk_{\k_1}(r)\Sk_{\k_2}^2(\theta)}\left( \frac{\beta_2}{
\cos^2\phi} +\frac{\beta_3 }{ \sin^2\phi} \right) ,\nonumber\\
L_2=J_{01}J_{12}-J_{03}J_{23}+k\, \k_2\Sk_{\k_2}(\theta)\cos\phi
-\frac{2\k_2\cos\phi}{\Tk_{\k_1}(r)}\left( \frac{\beta_1\Sk_{\k_2}(\theta)}{
\Ck_{\k_2}^2(\theta) } +\frac{\beta_3}{
\Sk_{\k_2}(\theta)\sin^2\phi} \right) ,\nonumber\\
L_3=J_{01}J_{13}+J_{02}J_{23}+k\, \k_2\Sk_{\k_2}(\theta)\sin\phi
-  \frac{2\k_2\sin\phi}{ \Tk_{\k_1}(r)}\left( \frac{\beta_1\Sk_{\k_2}(\theta)}{
\Ck_{\k_2}^2(\theta) } +\frac{ \beta_2 }{
\Sk_{\k_2}(\theta)\cos^2\phi}\right)   .
\label{hc}
\end{gather}

 The  superintegrability of {\it each} Hamiltonian ${\cal H}^{\rm GKC}_i={\cal T}+{\cal
U}^{\rm GKC}_i$ $(i=1,2,3)$  is determined by:
\begin{proposition}
{\rm (i)} The function $L_{i}$ \eqref{hc} Poisson commutes with   ${\cal
H}^{\rm GKC}_i$.

{\rm (ii)} The five  functions $\{L_i,I_{12},I_{23},I_{123},{\cal H}^{\rm GKC}_i\}$ are
functionally independent, thus
${\cal H}^{\rm GKC}_i$ is a  maximally superintegrable Hamiltonian.
\end{proposition}

\subsection[The Laplace-Runge-Lenz vector]{The Laplace--Runge--Lenz vector}

When another $\beta_j$ is taken equal to zero in a given potential ${\cal
U}^{\rm GKC}_i$ $(j\ne i)$,   the function $L_j$ is also  a constant  of
the motion. Therefore   when all the $\beta_j=0$, the three functions
(\ref{hc}) are constants of the motion for the GKC potential which reduces in this case to
the proper KC system. This is summed up in the following statements.

\begin{proposition}  Let one $\beta_j=0$ in the Hamiltonian ${\cal H}^{\rm GKC}_i={\cal
T}+{\cal U}^{\rm GKC}_i$ $(i=1,2,3)$  with $j\ne i$.

{\rm (i)} The  two functions $L_i,L_{j}$ Poisson commute with   ${\cal
H}^{\rm GKC}_i$.

{\rm (ii)} The   functions $\{I_{12},I_{23},I_{123},{\cal H}^{\rm GKC}_i\}$ together with
either $L_i$ or $L_j$ are functionally independent.
\end{proposition}

\begin{proposition} Let the three $\beta_i=0$, then:

{\rm (i)} The three GKC potentials reduce to its common $k$-term,  ${\cal U}^{\rm GKC}_i\equiv
{\cal U}^{\rm KC}=-k/\Tk_{\k_1}(r)$, which is the  (curved) KC potential on ${\mathbb
S}^3_{[\k_1]\k_2}$.

{\rm (ii)} The three
functions
 \begin{gather}
L_1=-J_{02}J_{12}-J_{03}J_{13}+k\, \k_2\Ck_{\k_2}(\theta),\nonumber\\
L_2=J_{01}J_{12}-J_{03}J_{23}+k\, \k_2\Sk_{\k_2}(\theta)\cos\phi,\nonumber\\
L_3=J_{01}J_{13}+J_{02}J_{23}+k\, \k_2\Sk_{\k_2}(\theta)\sin\phi,
\label{hhc}
\end{gather}
Poisson commutes with   ${\cal H}^{\rm KC}={\cal
T}+{\cal U}^{\rm KC}$ and these are the components of the Laplace--Runge--Lenz vector on
 ${\mathbb S}^3_{[\k_1]\k_2}$.

{\rm (iii)} The   functions $\{I_{12},I_{23},I_{123},{\cal H}^{\rm GKC}_i\}$ together with
any of the components $L_i$ are functionally independent.
\end{proposition}

On the other hand, equivalence amongst the
    Hamiltonians ${\cal H}^{\rm GKC}_i$ comes from their interpretation   on ${\mathbb
S}^3_{[\k_1]\k_2}$   that we proceed to study. We shall show that the three potentials
(\ref{hb}) are equivalent on the Riemannian spaces (take $i=3$), meanwhile we can distinguish
two different potentials on the spacetimes (take $i=1,3$). Thus we display in
Table~5 the corresponding non-equivalent GKC potentials together with the
additional constant of the motion (\ref{hc}).

\subsection{Description of the  GKC potential}

In Subsection 5.1 we have interpreted each of the $\beta_i$-terms appearing within the SW
potential  either as a non-central oscillator or  as a centrifugal barrier according to the
particular space under consideration. This, in turn, means that each   potential (\ref{hb})
is  a superposition of the   KC potential   with either oscillators or
centrifugal barriers. The latter interpretation arises directly by introducing the distances
$(x,y,z)$ (\ref{ggff}) and this holds simultaneously for the six spaces:
\begin{gather}
{\cal U}^{\rm GKC}_1=-\frac{k
}{\Tk_{\k_1}(r)}+\frac{\beta_2}{\Sk_{\k_1\k_2}^2(y)}
+\frac{\beta_3}{\Sk_{\k_1\k_2}^2(z) },\nonumber\\
{\cal U}^{\rm GKC}_2=-\frac{k
}{\Tk_{\k_1}(r)}+ \frac{\beta_1}{\Sk_{\k_1}^2 (x) }
+\frac{\beta_3}{\Sk_{\k_1\k_2}^2(z) } ,\nonumber\\
{\cal U}^{\rm GKC}_3=-\frac{k
}{\Tk_{\k_1}(r)}+ \frac{\beta_1}{\Sk_{\k_1}^2 (x) }+
\frac{\beta_2}{\Sk_{\k_1\k_2}^2(y)}.
\label{hd}
\end{gather}
 These expressions clearly show that some ${\cal H}^{\rm GKC}_i$ are  equivalent  according to
the value of $\k_2$, that is, the signature of the metric, so that we analyze  the two
possibilities separately.

\newpage

{\footnotesize
 \noindent
{\bf Table 5.} Maximally superintegrable generalized Kepler--Coulomb
potential ${\cal U}^{\rm
GKC}_i$, such that
${\cal H}^{\rm GKC}_i={\cal T}+{\cal U}^{\rm
GKC}_i$, and the additional
  constant of the motion $L_i$ to the set $\{I_{12},I_{23},I_{123}\}$ for
${\mathbb S}^3_{[\k_1]\k_2}$ with the same conventions given in Table~3 ($i=3$ for the
Riemannian spaces and $i=3,1$ for the spacetimes).
$$
\begin{array}{l}
\hline
\\[-7pt]
\multicolumn{1}{c}{\mbox {3D Riemannian spaces}} \\[3pt]
\hline
\\[-7pt]
\mbox {$\bullet$ Spherical space ${\bf S}^3$} \\[4pt]
\displaystyle{{\cal U}^{\rm
GKC}_3=-\frac{k
}{\tan r}
+\frac{1}{\sin^2 r}\left(\frac{\beta_1}{\cos^2\theta}+
\frac{\beta_2}{\sin^2\theta\cos^2\phi}  \right)}\\[8pt]
\displaystyle{ L_3=J_{01}J_{13}+J_{02}J_{23}+k \sin\theta\sin\phi
-  \frac{2 \sin\phi}{ \tan r}\left( \frac{\beta_1\sin\theta}{
\cos^2\theta } +\frac{ \beta_2 }{
\sin\theta\cos^2\phi}\right)   }\\[8pt]
\mbox {$\bullet$ Euclidean space  ${\bf E}^3$ } \\[4pt]
\displaystyle{ {\cal U}^{\rm
GKC}_3=-\frac{k
}{ r}
+\frac{1}{r^2}\left(\frac{\beta_1}{\cos^2\theta}+
\frac{\beta_2}{\sin^2\theta\cos^2\phi}\right)}\\[8pt]
\displaystyle{ L_3=J_{01}J_{13}+J_{02}J_{23}+k \sin\theta\sin\phi
-  \frac{2 \sin\phi}{  r}\left( \frac{\beta_1\sin\theta}{
\cos^2\theta } +\frac{ \beta_2 }{
\sin\theta\cos^2\phi}\right)    }\\[8pt]
\mbox {$\bullet$ Hyperbolic space ${\bf H}^3$  }\\[4pt]
\displaystyle{ {\cal U}^{\rm
GKC}_3=-\frac{k
}{\tanh r}
+\frac{1}{\sinh^2 r}\left(\frac{\beta_1}{\cos^2\theta}+
\frac{\beta_2}{\sin^2\theta\cos^2\phi} \right)}\\[8pt]
\displaystyle{ L_3=J_{01}J_{13}+J_{02}J_{23}+k \sin\theta\sin\phi
-  \frac{2 \sin\phi}{ \tanh r}\left( \frac{\beta_1\sin\theta}{
\cos^2\theta } +\frac{ \beta_2 }{
\sin\theta\cos^2\phi}\right)    }\\[8pt]
 \hline
\\[-7pt]
\multicolumn{1}{c}{\mbox {$(2+1)$D
 Relativistic spacetimes
}} \\[3pt]
\hline
\\[-7pt]
\mbox {$\bullet$ Anti-de Sitter
spacetime $ {\bf AdS}^{2+1}$   } \\[4pt]
\displaystyle{ {\cal U}^{\rm
GKC}_3=-\frac{k
}{\tan r}
+\frac{1}{\sin^2 r}\left(\frac{\beta_1}{\cosh^2\theta}+
\frac{\beta_2}{\sinh^2\theta\cos^2\phi}  \right)
}\\[8pt]
\displaystyle{  L_3=J_{01}J_{13}+J_{02}J_{23}-k \sinh\theta\sin\phi
+  \frac{2 \sin\phi}{ \tan r}\left( \frac{\beta_1\sinh\theta}{
\cosh^2\theta } +\frac{ \beta_2 }{
\sinh\theta\cos^2\phi}\right)    }\\[8pt]
\displaystyle{ {\cal U}^{\rm
GKC}_1=-\frac{k
}{\tan r}
+\frac{1}{\sin^2 r\sinh^2\theta}\left(\frac{\beta_2}{\cos^2\phi}+
\frac{\beta_3}{ \sin^2\phi}  \right)}\\[8pt]
\displaystyle{L_1=-J_{02}J_{12}-J_{03}J_{13}-k \cosh\theta
+\frac{2 \cosh\theta}{\tan r\sinh^2\theta}\left( \frac{\beta_2}{
\cos^2\phi} +\frac{\beta_3 }{
\sin^2\phi}
\right) }\\[8pt]
\mbox {$\bullet$ Minkowskian spacetime ${\bf
M}^{2+1}$  } \\[4pt]
\displaystyle{ {\cal U}^{\rm
GKC}_3=-\frac{k
}{ r}
+\frac{1}{r^2}\left(\frac{\beta_1}{\cosh^2\theta}+
\frac{\beta_2}{\sinh^2\theta\cos^2\phi} \right) }\\[8pt]
\displaystyle{  L_3=J_{01}J_{13}+J_{02}J_{23}-k \sinh\theta\sin\phi
+  \frac{2 \sin\phi}{  r}\left( \frac{\beta_1\sinh\theta}{
\cosh^2\theta } +\frac{ \beta_2 }{
\sinh\theta\cos^2\phi}\right)    }
\\[8pt]
\displaystyle{   {\cal U}^{\rm
GKC}_1=-\frac{k
}{ r}
+\frac{1}{ r^2\sinh^2\theta}\left(\frac{\beta_2}{\cos^2\phi}+
\frac{\beta_3}{ \sin^2\phi}  \right)}\\[8pt]
\displaystyle{L_1=-J_{02}J_{12}-J_{03}J_{13}-k \cosh\theta
+\frac{2 \cosh\theta}{ r\sinh^2\theta}\left( \frac{\beta_2}{
\cos^2\phi} +\frac{\beta_3 }{
\sin^2\phi}
\right) }\\[8pt]
\mbox {$\bullet$ De Sitter
spacetime
 $ {\bf dS}^{2+1}$  }\\[4pt]
\displaystyle{ {\cal U}^{\rm
GKC}_3=-\frac{k
}{\tanh r}
+\frac{1}{\sinh^2 r}\left(\frac{\beta_1}{\cosh^2\theta}+
\frac{\beta_2}{\sinh^2\theta\cos^2\phi} \right)   }\\[8pt]
\displaystyle{  L_3=J_{01}J_{13}+J_{02}J_{23}-k \sinh\theta\sin\phi
+  \frac{2 \sin\phi}{ \tanh r}\left( \frac{\beta_1\sinh\theta}{
\cosh^2\theta } +\frac{ \beta_2 }{
\sinh\theta\cos^2\phi}\right)    }\\[8pt]
\displaystyle{ {\cal U}^{\rm
GKC}_1=-\frac{k
}{\tanh r}
+\frac{1}{\sinh^2 r\sinh^2\theta}\left(\frac{\beta_2}{\cos^2\phi}+
\frac{\beta_3}{ \sin^2\phi}  \right) }\\[8pt]
\displaystyle{L_1=-J_{02}J_{12}-J_{03}J_{13}-k \cosh\theta
+\frac{2 \cosh\theta}{\tanh r\sinh^2\theta}\left( \frac{\beta_2}{
\cos^2\phi} +\frac{\beta_3 }{
\sin^2\phi}
\right) }\\[8pt]
\hline
\end{array}
$$
}

\subsubsection{Riemannian spaces}

When $\k_2=+1$ the three distances $(x,y,z)$ are completely equivalent, and their ``label" in
the trigonometric functions is always $\k_1$ (recall that in this case both $\theta$ and
$\phi$ are ordinary angles). Hence the three Hamiltonians ${\cal H}^{\rm GKC}_i$ are
also equivalent and we only consider a unique potential, say
${\cal U}^{\rm GKC}_3$ with constant of the motion $L_3$. On the spherical space ${\bf S}^3$
both
$\beta_1$,
$\beta_2$ terms can alternatively be expressed as non-central oscillators as commented in
Subsection 5.1.1, meanwhile on
${\bf E}^3$ and ${\bf H}^3$ these only can be interpreted as centrifugal barriers. In this
way we find the following expressions for each  space:
\begin{alignat*}{5}
& {\bf S}^3:\quad && {\cal U}^{\rm GKC}_3&=& -\frac{k
}{\tan r}+ \frac{\beta_1}{\sin^2 x }+ \frac{\beta_2}{\sin^2 y}
= -\frac{k }{\tan r}+  {\beta_1}{\tan^2 r_1 }+ {\beta_2}{\tan^2 r_2}+\beta_1+\beta_2;& \\
&{\bf E}^3:\quad && {\cal U}^{\rm GKC}_3 &= & -\frac{k}{r}+ \frac{\beta_1}{ x^2 }+
\frac{\beta_2}{ y^2};& \\
& {\bf H}^3:\quad && {\cal U}^{\rm GKC}_3 &=& -\frac{k
}{\tanh r}+ \frac{\beta_1}{\sinh^2 x }+ \frac{\beta_2}{\sinh^2 y}.&
\end{alignat*}
When all the $\beta_i=0$ we obtain the   components of the Laplace--Runge--Lenz vector
(\ref{hhc}) for the three Riemannian spaces:
 \begin{gather*}
L_1=-J_{02}J_{12}-J_{03}J_{13}+k\cos \theta,\\
L_2=J_{01}J_{12}-J_{03}J_{23}+k \sin\theta\cos\phi  ,\\
L_3=J_{01}J_{13}+J_{02}J_{23}+k \sin\theta\sin\phi,
\end{gather*}
where the difference for each particular space comes from the translations $J_{0i}$
(\ref{ee}) that do depend on the curvature $\k_1$.

\subsubsection{Relativistic spacetimes}

On the contrary, if $\k_2=-1$ (in units $c=1$), only the two space-like distances $y$ and $z$
are~equi\-valent while $x$ is a~time-like distance ($\phi$ is also an angle for the three
spacetimes but $\theta$ is a~rapidity).  Thus ${\cal U}^{\rm GKC}_2\simeq {\cal U}^{\rm
GKC}_3$ containing a time-like centrifugal barrier and another space-li\-ke one, while
  ${\cal U}^{\rm GKC}_1$ defines a different potential with two space-like centrifugal
barriers for the three spacetimes. By taking into account the results given in Subsection 5.1,
these potentials~show different superpositions of the KC potential with
non-central harmonic oscillators and cent\-ri\-fu\-gal barriers on  ${\bf AdS}^{2+1}$ and ${\bf
dS}^{2+1}$.   The explicit expressions on each spacetime   turn out~to~be
\begin{alignat*}{5}
&{\bf AdS}^{2+1}:\quad && {\cal U}^{\rm GKC}_3 &=& -\frac{k
}{\tan r}+ \frac{\beta_1}{\sin^2 x }+ \frac{\beta_2}{\sinh^2 y} = -\frac{k
}{\tan r}+  {\beta_1}{\tan^2 r_1 }+
\frac{\beta_2}{\sinh^2 y}+\beta_1,&\\
& && {\cal U}^{\rm GKC}_1 &= & -\frac{k }{\tan r}+ \frac{\beta_2}{\sinh^2 y }+
\frac{\beta_3}{\sinh^2 z}  ; &\\
&{\bf M}^{2+1}:&& {\cal U}^{\rm GKC}_3 &=&  -\frac{k}{r}+ \frac{\beta_1}{ x^2 }+
\frac{\beta_2}{ y^2}, &\\
&& &  {\cal U}^{\rm GKC}_1 &= & -\frac{k }{r}+ \frac{\beta_2}{ y^2 }+
\frac{\beta_3}{ z^2};&\\
&{\bf dS}^{2+1}:&& {\cal U}^{\rm GKC}_3 &=&  -\frac{k }{\tanh r}+ \frac{\beta_1}{\sinh^2 x }+
\frac{\beta_2}{\sin^2 y} = -\frac{k }{\tanh r}+ \frac{\beta_1}{\sinh^2 x}+  {\beta_2}{\tan^2 r_2 }+  \beta_2, &\\
& && {\cal U}^{\rm GKC}_1 &=&  -\frac{k }{\tanh r}+ \frac{\beta_2}{\sin^2 y}+
\frac{\beta_3}{\sin^2 z}  &\\
&&& &=& -\frac{k
}{\tanh r}+  {\beta_2}{\tan^2 r_2 }+  {\beta_3}{\tan^2 r_3 }+
 \beta_2+\beta_3. &
\end{alignat*}
The   components of the Laplace--Runge--Lenz vector (\ref{hhc}) (for $\beta_i=0$)
written in terms of the kinematical generators (\ref{be}) are
 \begin{gather*}
L_1=-P_1 K_1-P_2K_2-k \cosh \theta,\qquad
L_2=P_0K_1-P_2 J-k \sinh \theta\cos\phi,\\
L_3=P_0K_2+P_1J -k \sinh\theta\sin\phi.
\end{gather*}

\section{Concluding remarks}

We have achieved the generalization of the 3D Euclidean superintegrable family (\ref{aa}) as well as
the maximally superintegrable SW (\ref{ab}) and GKC (\ref{ac}) potentials to the space ${\mathbb
S}^3_{[\k_1]\k_2}$ by applying a unified approach which makes use of a built-in scheme of
contractions. Furthermore the   results so obtained have been described and interpreted on each
particular space  and have also been displayed along the paper in tabular form. Thus we have
explicitly shown that (maxi\-mal) superintegrability  is preserved for any value of the curvature and
for either a Riemannian or Lorentzian metric. Notice that on the complex sphere (see
e.g.~\cite{21}) and on the ambient space~$\mathbb R^4$ these two  maximally superintegrable
Hamiltonians read
\[
{\cal H}^{\rm
SW}=\sum_{\mu=0}^3\left(
\frac 12\, p_\mu^2 +\frac{\beta_\mu}{x_\mu^2}\right)-\beta_0 ,\qquad
{\cal H}^{\rm
GKC}_3=\frac 12\sum_{\mu=0}^3
 p_\mu^2  -\frac{k\,x_0}{
\sqrt{x_1^2+x_2^2+x_3^2}}    +\frac{\beta_1}{x_1^2}    +\frac{\beta_2}{x_2^2} ,
\]
where $ \sum\limits_{\mu=0}^3 x_\mu^2=1$. Therefore the Hamiltonians here studied
can be regarded as different real forms coming from  known complex superintegrable systems through
graded contractions, that is, by introducing the parameters $\k_1$ and $\k_2$.

As far as the superintegrable potential ${\cal U}$ (\ref{fa}) is concerned,
we recall that in this 3D case, we have one   constant of the motion (besides de
Hamiltonian) more than  the two ones that ensure its complete integrability,
but one integral less than the four ones that determine   maximal superintegrability.
By taking into account the former point of view one may claim that ${\cal U}$ is   {\it minimally}
(or weak) superintegrable, while from the latter, ${\cal U}$ would be  {\it
quasi-maximally} superintegrable.  Our opinion is that when the corresponding
  Hamiltonian ${\cal H}={\cal T}+{\cal U}$ is constructed on the  $N$D spaces ${\mathbb
S}^N_{[\k_1]\k_2}$,   each of the $N(N-1)$  generators $J_{ij}$ $(i,j=1,\dots,N;\ i<j)$ of the
(Lorentz) rotation subalgebra $so_{\k_2}(N)$ would provide a constant of the motion $I_{ij}$
of the type~(\ref{int}). Next, by following~\cite{VulpiLett,CRMVulpi}, two subsets of  $N-1$
constants of the motion, $Q_{(l)}$ and
$Q^{(l)}$, should be deduced from the initial set  of $N(N-1)$ integrals
as:
\begin{gather}
 Q^{(l)} =\sum_{i,j=1}^l
I_{ij} ,\qquad
 Q_{(l)} =\sum_{i,j=N-l+1}^N\!\!
I_{ij}     ,\qquad l=2,\dots,N,
\label{sax}
\end{gather}
where $ Q^{(N)}\equiv Q_{(N)}$. In this way the complete integrability of  ${\cal H}$ would be
characterized by either the $N$ constants of the motion $\{ Q^{(l)} ,{\cal H} \}$ or by $\{ Q_{(l)}
,{\cal H} \}$. The quasi-maximal superintegrability would be provided by the $2N-2$ functions
\[
\{ Q^{(2)}, Q^{(3)},\dots,  Q^{(N)}\equiv Q_{(N)},\dots,Q_{(3)},  Q_{(2)}, \cal H\}.
\]

The corresponding SW potential on ${\mathbb S}^N_{[\k_1]\k_2}$
 would be obtained by taking the same ${\cal F}(r)$ as in~(\ref{ga}) and the
remaining constant of the motion would come from one of the translation generators~$J_{0i}$ in the
form $I_{0i}$ (\ref{inta}). Likewise, a set of $N$ GKC potentials could be constructed by starting
from the radial function (\ref{ha}) and then taking $N-1$ centrifugal terms for each of them as in
(\ref{hb}); in this case  the additional constant of the motion $L_i$  would be of the form~(\ref{intb}).

We stress that this scheme of the possible $N$D generalization of all the 3D results here presented
(currently in progress) relies on the fact that the potential
${\cal U}$ can be endowed with a~coalgebra symmetry~\cite{5}. This indeed allowed us to obtain
the integrals (\ref{sax}) for the $N$D SW system on the three Riemannian spaces
in~\cite{VulpiLett,CRMVulpi} by starting from the quantum deformation of the Euclidean SW system
introduced in~\cite{1,2}. Furthermore, quantum deformations have been shown~\cite{plb,czec} to give
rise to Riemannian and relativistic spaces of   non-constant curvature on which SW- and KC-type
potentials can be considered~\cite{jpa2D}.

\subsection*{Acknowledgements}

 This work was partially supported  by the Ministerio de Educaci\'on y
Ciencia   (Spain, Project FIS2004-07913) and  by the Junta de Castilla y
Le\'on   (Spain, Projects  BU04/03 and VA013C05).

\LastPageEnding

\end{document}